
\magnification\magstep1

\font\smc=cmcsc10                 
\font\eightrm=cmr8                
\font\eightmi=cmmi8               

\font\tenbbb=msbm10 \font\sevenbbb=msbm7
\newfam\bbbfam
\textfont\bbbfam=\tenbbb \scriptfont\bbbfam=\sevenbbb
\def\Bbb{\fam\bbbfam}             

\font\tengoth=eufm10 \font\sevengoth=eufm7
\newfam\gothfam
\textfont\gothfam=\tengoth \scriptfont\gothfam=\sevengoth
\def\goth{\fam\gothfam}           

\font\tenbm=cmmib10 \font\sevenbm=cmmib10 at 7pt
\newfam\bmfam
\textfont\bmfam=\tenbm \scriptfont\bmfam=\sevenbm
\def\bm{\fam\bmfam}               

\mathchardef\alb="7A0B            
\mathchardef\sgb="7A1B            

\def\opname#1{\mathop{\rm#1}\nolimits} 

\def\a{\alpha}                    
\def\abs#1{\vert#1\vert}          
\def\Ad{\opname{Ad}}              
\def\adv{{\rm adv}}               
\def\Ag{{\goth A}}                
\def\baja{{\mathord\downarrow}}   
\def\braCket#1#2#3{\langle#1\mathbin\vert
                   #2\mathbin\vert#3\rangle}     
\def\braket#1#2{\langle#1\mathbin\vert#2\rangle} 
\def\bbraket#1#2{\langle\!\langle#1\mathbin\vert
                 #2\rangle\!\rangle}  
\def\C{{\Bbb C}}                  
\def\cite#1{$^{#1}$}              
\def\comm#1#2{\lbrack#1,#2\rbrack} 
\def\ddto#1{{d\over{d#1}}\biggr\vert_{#1=0}} 
\def\det{\opname{det}}            
\def\detc{\opname{det_\C}}        
\def\dm{\dot\mu}                  
\def\Dom{\opname{Dom}}            
\def\End{\opname{End}}            
\def\Endr{\End_\R}                
\def\eps{\epsilon}                
\def\eq#1{{\rm(#1)}}              
\def\F{{\cal F}}                  
\def\ga{\gamma}                   
\def\Ga{\Gamma}                   
\def\H{{\cal H}}                  
\def\Hat#1{\widehat{#1}}          
\def\half{{1\over2}}              
\def\ind{\opname{ind}}            
\def\J{{\cal J}}                  
\def\kb{{\bm k}}                  
\def\ket#1{\vert#1\rangle}            
\def\La{\Lambda}                  
\def\la{\lambda}                  
\def\N{{\Bbb N}}                  
\def\norm#1{\Vert#1\Vert}         
\def\ol{{\goth o}}                
\def\Onda#1{\widetilde{#1}}       
\def\pb{{\bm p}}                  
\def\Pf{\opname{Pf}}              
\def\Proof{\noindent{\sl Proof\/}}   
\def\Q{{\Bbb Q}}                  
\def\R{{\Bbb R}}                  
\def\ret{{\rm ret}}               
\def\sepword#1{\qquad\hbox{#1}\quad} 
\def\set#1{\{\,#1\,\}}            
\def\shalf{{\scriptstyle{1\over2}}}  
\def\Sk{{\rm Sk}}                 
\def\sube{{\mathord\uparrow}}     
\def\thalf{{\textstyle{1\over2}}} 
\def\tihalf{{\textstyle{i\over2}}}    
\def\tquarter{{\textstyle{1\over4}}}  
\def\Tr{\opname{Tr}}              
\def\vac{{\ket 0}}                
\def\vacpersamp{\<0_{\rm in},0_{\rm out}>} 
\def\vacout{\ket{0_{\rm out}}}    
\def\w{\wedge}                    
\def\whack#1{\not{\!\!#1}}        
\def\wick:#1:{\mathopen:#1\mathclose:} 
\def\wrt{with respect to}         
\def\x{\times}                    
\def\xb{{\bm x}}                  
\def\Z{{\Bbb Z}}                  
\def\1{{\'\i}}                    
\def\7{\dagger}                   
\def\.{\cdot}                     
\def\:{\colon}                    
\def\<#1,#2>{\langle#1\mathbin\vert#2\rangle} 

\def\qed{\allowbreak\qquad\null
         \nobreak\hfill\square}   

\def\square{\vbox{\hrule
             \hbox{\vrule height 5.2pt
              \hskip 5.2pt
              \vrule}\hrule}}     

\outer\def\section#1. #2\par{
      \bigskip\bigskip
      \message{#1. #2}%
      \leftline{\bf#1. #2}\nobreak
      \smallskip\noindent}

\outer\def\subsection#1. #2\par{
      \bigskip \vskip\parskip
      \message{\string\S\space#1.}%
      \leftline{\rm#1.\enspace\it#2}\nobreak
      \smallskip\noindent}

\def\refno#1. #2, #3\par{\smallskip 
      \indent\hang\llap{$^{#1}\,$}%
      {\rm#2, }#3\par}

\newbox\bigstrutbox               
\setbox\bigstrutbox=\hbox to0pt{} 
    \ht\bigstrutbox=12pt
    \dp\bigstrutbox= 4pt
\def\Strut{\copy\bigstrutbox}


\def\curraddr{{\rm a)}}
\def\ShaleStine{1}
\def\Pressley{2}
\def\Mickelsson{3}
\def\Connes{4}
\def\Araki{5}
\def\Bourbaki{6}
\def\Emch{7}
\def\Taylor{8}
\def\Segalcwave{9}
\def\Ruijsenaars{10}
\def\Lawson{11}
\def\Sirius{12}
\def\CareyPalmer{13}
\def\Palmer{14}
\def\Popov{15}
\def\Borthwick{16}
\def\GrosseMad{17}
\def\Mickelssonbis{18}
\def\Maderner{19}
\def\Weinberg{20}
\def\Bongaarts{21}
\def\Saunders{22}
\def\Thaller{23}
\def\Careyob{24}
\def\Matsui{25}
\def\AGaume{26}
\def\Ruijsenaarsbis{27}
\def\Aldebaran{28}
\def\Scharf{29}
\def\Feynman{30}
\def\BogolShir{31}
\def\Dittrich{32}


\centerline{\bf QED in external fields
                from the spin representation}

\bigskip

\centerline{\smc Jos\'e M. Gracia-Bond{\'\i}a* and
                 Joseph C. V\'arilly\dag$^{\rm a)}$}

\bigskip

\centerline{*\it Departamento de F{\'\i}sica Te\'orica,
         Universidad Aut\'onoma de Madrid, 28049 Madrid, Spain}
\smallskip
\centerline{and}
\smallskip
\centerline{*\it Departamento de F{\'\i}sica Te\'orica,
                 Universidad de Zaragoza, 50009 Zaragoza, Spain}
\medskip
\centerline{\dag\it Departamento de Matem\'aticas, Centro de
                    Investigaci\'on y Estudios Avanzados}
\centerline{\it     Instituto Polit\'ecnico Nacional,
                    07000 M\'exico DF, M\'exico}

\bigskip

\centerline{\smc Abstract}
\smallskip

{\narrower\noindent
\eightrm\baselineskip=9.5pt
Systematic use of the infinite-dimensional spin representation
simplifies and rigo\-rizes several questions in Quantum Field
Theory. This representation permutes ``Gaussian'' elements in the
fermion Fock space, and is necessarily projective: we compute
its cocycle at the group level, and obtain Schwinger terms and
anomalies from infinitesimal versions of this cocycle.
Quantization, in this framework, depends on the choice of the
``right'' complex structure on the space of solutions of the
Dirac equation. We show how the spin representation allows one
to compute exactly the {\eightmi S}-matrix for fermions in an
external field; the cocycle yields a causality condition needed
to determine the phase.
\par\smallskip}



\section I. Introduction

It should have been clear since Shale and Stinespring's seminal
paper\cite{\ShaleStine} that the des\-cription of fermions
coupled to external classical fields, such as gauge fields,
reduces to a problem in representation theory of the infinite
dimensional orthogonal group. That paper was couched in general
theoretical terms and no explicit calculation was made. The spin
(and pin) representation came to the fore in Quantum Field
Theory again as the cornerstone of the remarkable books by
Pressley and Segal,\cite{\Pressley} dealing with loop groups
---not unrelated to the subjects of string theory, Kac--Moody
algebras and integrable systems--- and
Mickelsson,\cite{\Mickelsson} dealing with current algebras.
However, the spin representation is not calculated in all
generality in these books.

   We give the pin representation of the infinite-dimensional
orthogonal group {\it\`a la\/} Pressley and Segal, in full
detail, and we derive from it, among other things, the
scattering matrix in closed form and the Feynman rules. For
simplicity, we consider only fermions coupled to external
electromagnetic fields in Minkowski space (the external field
problem in QED), although the realm of applicability of the spin
representation is much wider. We hope to convince the reader
that ours is a very economical approach to linear Quantum Field
Theory, and that there is nothing in this branch of quantum
electrodynamics that cannot be traced back to the
representation. Applications of the pin representation to
nonlinear field theories will be examined in a separate paper.
We have taken pains to give a rigorous treatment; except in the
last Section, smeared field operators are used throughout.

   Section~II reviews the algebraic theory of infinite
dimensional vector spaces with a symmetric form, dealing with
the respective complex structures and polarizations. Section~III
brings in the fermion Fock space and canonical anticommutation
relations. The main tool is a series expansion of the general
``Gaussian'' element of this space, whose coefficients are
Pfaffians of skewsymmetric operators. This will be needed in the
following Section~IV, where we construct the pin representation
for the orthogonal group; actually, the group acting on Fock
space is an extension by $U(1)$.

   The pin representation immediately proves its worth in
yielding the quantization prescription by means of its
infinitesimal version. We show how the cocycle of the pin
representation gives the general anomaly and the anomalous
commutators (Schwinger terms) for linear fermion fields. All
this is dealt with in Section~V. Our formulas concerning the
anomaly appear to be new. In Section~VI we rewrite the
representation in terms of field operators on Fock space.

   In Section~VII, after a discussion of the space of solutions
of the Dirac equation in the framework of Section~II, an
all-important step is taken, to wit, the choice of complex
structure, which is determined by the nature of the vacuum in
quantum electrodynamics. It turns out ---somewhat
mysteriously--- that the correct complex structure is closely
related to the phase of the Dirac operator, which plays a
prominent r\^ole in Connes's\cite{\Connes} noncommutative
differential geometry. We then complete a careful translation
between the language of group representation theory and that of
quantum electrodynamics. After treating charge sectors, we
develop the exact expression of the $S$-matrix for charged
fermions in an external field, and we reexpress the Schwinger
terms and the general anomaly formula directly in terms of the
scattering operator for the Dirac equation.

   In Section~VIII, we derive in a completely rigorous manner
the Feynman rules for electrodynamics in external fields,
within the validity conditions of the classical perturbation
expansion. Section~IX briefly deals with vacuum polarization.

   Throughout, units are so taken that $c = 1$ and $\hbar = 1$.

\section II. The infinite-dimensional orthogonal group:
            algebraic aspects

The treatment of the symmetries of the fermion field which we
develop here starts from the observation that the space of
solutions of the Dirac equation with an external potential is a
{\it real\/} vector space with a distinguished symmetric form.

\subsection II.1. Orthogonal complex structures

We start with a real vector space $V$ and a symmetric bilinear
form~$d$, given {\it a priori}. We lose nothing by supposing $V$
to be complete in the metric induced by~$d$, so we take $(V,d)$
to be a real Hilbert space, either infinite-dimensional or of
finite even dimension.

   An orthogonal {\it complex structure\/}~$J$ is a real-linear
operator on~$V$ satisfying:
$$
J^2 = -I,  \qquad\hbox{and}\qquad  d(Ju,Jv) = d(u,v)
 \quad\hbox{for } u,v\in V.
$$
Now, regarding $V$ as a complex vector space via the rule
$(\a + i\beta)v := \a v + \beta Jv$ for $\a,\beta$ real, the
hermitian form
$$
\<u,v>_{\scriptscriptstyle J} := d(u,v) + id(Ju,v)
$$
makes $(V,d,J)$ a complex Hilbert space.

   The orthogonal group $O(V)$ is
$\{\,g\in GL_\R(V) : d(gu,gv) = d(u,v)$, for all $u,v\in V\,\}$.
Note that $g$ is orthogonal iff $g^tg = I$ where the transpose
is \wrt~$d$.

   The set $\J(V)$ of orthogonal complex structures on~$V$ can
be seen as a subset of the orthogonal Lie algebra
$\ol(V) = \set{X\:V \to V\ \hbox{real-linear}:
  d(\.,X\.) + d(X\.,\.) = 0}$ of~$O(V)$, preserved by the
adjoint action $J' \mapsto gJ'g^{-1}$ (with $J' \in \J(V)$) of
the orthogonal group.

   We select a particular complex structure called $J$ and
decompose elements of $O(V)$ as $g = p_g + q_g$ where $p_g$,
$q_g$ are its linear and antilinear parts:
$p_g := \thalf(g - JgJ)$, $q_g := \thalf(g + JgJ)$. We find that
$p_{g^{-1}} = \thalf(g^{-1} - Jg^{-1}J) = \thalf(g^t - Jg^tJ)
 = p_g^t$, whereas $q_{g^{-1}} = \thalf(g^{-1} + Jg^{-1}J)
 = \thalf(g^t + Jg^tJ) = q_g^t$. Linear and antilinear parts of
$gg^{-1} = g^{-1}g = I$ yield, for $g \in O(V)$:
$$
p_g p_g^t + q_g q_g^t = p_g^t p_g + q_g^t q_g = I,  \quad
p_g q_g^t = - q_g p_g^t, \quad  p_g^t q_g = - q_g^t p_g.
\eqno (2.1)
$$

   The complexification $V_\C = V \oplus iV$ is a complex
Hilbert space under the positive-definite hermitian form:
$$
\bbraket{w_1}{w_2} := 2 d(w_1^*,w_2).
$$
Writing $P_J := \thalf(I - iJ)$, $W_0 := P_J V_\C = P_J V$ is a
Hilbert subspace of~$V_\C$, satisfying
$V_\C = W_0 \oplus W_0^*$, and
$$
\bbraket{P_J u}{P_J v} = \<u,v>,  \qquad
 \bbraket{P_{-J}u}{P_{-J}v} = \<v,u>  \sepword{for} u,v \in V.
$$
Moreover, $W_0$ is isotropic for~$d$, i.e.,
$d(u - iJu, v - iJv) = 0$ for $u,v \in V$.

   A (complex) {\it polarization\/} for~$d$ is any complex
subspace $W \leq V_\C$ which is isotropic for~$d$ and satisfies
$W \cap W^* = 0$, $W \oplus W^* = V_\C$. If $w \in W$, then
$w = u - iv$ for unique elements $u,v \in V$, and
$J_W\: u \mapsto v$ is real-linear; thus
$W = \set{u - iJ_Wu : u \in V}$. Now $\Re\,d(w_1,w_2) = 0$ for
$w_1,w_2 \in W$ shows that $J_W$ is orthogonal, and
$\Im\,d(w_1,w_2) = 0$ gives $J_W^2 = -I$. Conversely, if
$J' \in \J(V)$, then $W' := \set{u - iJ'u : u \in V}$ is a
polarization for~$d$. The correspondence $W \mapsto J_W$
intertwines the adjoint action of~$O(V)$ on~$\J(V)$ and its
action $W \mapsto gW$ on the set of polarizations.

   If $W_1,W_2$ are two polarizations for~$d$, we can find a
unitary transformation $U\: W_1 \to W_2$; if $C$ denotes complex
conjugation in~$V_\C$, then $g = U \oplus CUC$ is a unitary
operator on $V_\C$ commuting with~$C$, so $g \in O(V)$ and
$gW_1 = W_2$. Thus $O(V)$ acts transitively on polarizations, and
hence also on~$\J(V)$. The isotropy subgroup of~$J$ in $\J(V)$
is $U_J(V)$, the unitary group of the Hilbert space $(V,d,J)$.

\subsection II.2. The restricted orthogonal group

We define the {\it restricted orthogonal group\/} $O'(V)$ to be
the subgroup of~$O(V)$ consisting of those $g$ for which
$\comm Jg$ is a Hilbert--Schmidt operator, or equivalently, for
which $q_g$ is Hilbert--Schmidt. Similarly, we restrict the set
of complex structures by introducing
$\J'(V) := \{\,J' \in \J(V) : J - J'$ is Hilbert--Schmidt$\,\}$.
Also, we call ``restricted polarizations'' those $W$ for which
$J - J_W$ is Hilbert--Schmidt; these form the orbit of $W_0$
under $O'(V)$. Since $U_J(V) = \set{g \in O(V) : \comm Jg = 0}$,
it is again the isotropy subgroup of $J$ or~$W_0$ under the
respective actions of~$O'(V)$.

   In the finite-dimensional case, we thus have
$\J'(V) \approx O(2n)/U(n)$, which has two connected components,
one of which is $SO(2n)/U(n)$. It can be shown that $J_W$ is in
the same connected component as~$J$ iff $\dim(W \cap W_0^*)$ is
even. In the infinite-dimensional case, the same results hold
true\cite{\Araki}: $\J'(V)$ has two connected components, and
the component of any $J_W \in \J'(V)$ is determined by the
parity of $\dim(W \cap W_0^*)$. Likewise, $O'(V)$ has two
components: $g$ lies in the neutral component iff
$\dim(g W_0 \cap W_0^*)$ is even. We shall denote by~$SO'(V)$
the component of the identity of~$O'(V)$.

   To see that $\dim(W \cap W_0^*)$ is finite, we note that
$B := \thalf(I - J_WJ)$ is a Fredholm operator on the complex
Hilbert space $V_\C$, since
$B^\7 B - I = B B^\7 - I = \tquarter(J_W - J)^2$ is traceclass.
Moreover,
$$
\eqalignno{
\ker B = \ker B^\7
&= \set{z \in V_\C : Jz = -J_Wz}  \cr
&= \set{z \in V_\C : \thalf(I \mp iJ)z = \thalf(I \pm iJ_W)z} \cr
&= (W \cap W_0^*) \oplus (W^* \cap W_0).
  & (2.2) \cr}
$$
In particular, $B$ has index:
$\dim(\ker B) - \dim(\ker B^\7) = 0$. Since
$W^* \cap W_0 = C(W \cap W_0^*)$, we have
$\dim(W \cap W_0^*) = \thalf \dim(\ker B)$, which is finite. Note
that $B = g p_g^t$, from which
$\dim_\C(\ker p_g) = \dim_\C(\ker p_g^t) = \thalf \dim(\ker B)$;
so $p_g$ is a Fredholm operator of index zero.

   Note also that $W \cap W_0^* = \{0\}$ iff $\thalf(I - J_WJ)$
is invertible iff $\norm{J_W - J} < 2$. For such~$W$, we can
write $w = z_1 + z_2^*$ with $z_j = u_j - iJu_j \in W_0$ for
$j = 1,2$. This defines a real-linear operator $T_W$ on~$V$ by
$T_W(u_1) := u_2$. By examining $iw = iz_1 - (iz_2)^*$, we find
that $T_WJ = - JT_W$, i.e., $T_W$ is antilinear. Moreover,
$$
0 = \thalf\Re\, d(w,w')
 = \thalf\Re\, (d(z_1,{z'_2}^*) + d(z'_1,z_2^*))
 = d(u_1,T_Wu'_1) + d(T_Wu_1,u'_1),
$$
so that $T_W$ is {\it skewsymmetric}. Since
$w = z_1 + z_2^* = (I + T_W)u_1 - iJ(I - T_W)u_1 = (I + T_W)z_1$,
we find that $J_W$ and $T_W$ are related by a pair of Cayley
transformations:
$$
J_W = J(I - T_W)(I + T_W)^{-1},  \qquad
 T_W = (J - J_W)(J + J_W)^{-1},
$$
using that $\ker(J_W + J) = \{0\}$ whenever
$W \cap W_0^* = \{0\}$ by~\eq{2.2}; hence $T_W$ is a
Hilbert--Schmidt operator. In synthesis, $T_W \in \Sk(V)$, where
$\Sk(V)$ denotes the vector space of {\it antilinear
skewsymmetric Hilbert--Schmidt operators\/} on~$V$.

\smallskip

   When $p_g$ is invertible, we can define
$T_g := q_g p_g^{-1}$. We have $T_g \in \Sk(V)$, since it equals
$T_{gW_0}$, as is readily checked; and from~\eq{2.1} we see
that  $p_g^t(I - T_g^2)p_g = I$. This shows that we may regard
the pair of operators $(p_g,T_g)$ as a parametrization of
$SO'_*(V) := \set{g \in SO'(V) : p_g^{-1}\rm\ exists}$.
While this is not a subgroup of $SO'(V)$, it is an open
neighbourhood of the identity in the topology of~$SO'(V)$
induced by the norm
$g \mapsto \norm{g}_{\rm op} + \norm{\comm Jg}_{\rm HS}$. Any
element of $O'(V)$ not in $SO'_*(V)$ satisfies
$\dim_\C(\ker p_g) = n > 0$ and one can find $r \in O'(V)$ which
is a product of $n$~reflections fixing $(\ker p_g^t)^\perp$, for
which $rg \in SO'_*(V)$. We will therefore devote most attention
to the case $g \in SO'_*(V)$.

   A few formulae for $p_g$ and $T_g$ will be very useful later.
Let us abbreviate $\Hat T_g := T_{g^{-1}}$. From the antilinear
part of the equation
$I = gg^{-1} = (I + T_g)p_g(I + \Hat T_g)p_g^{-1}$ we obtain
$0 = T_gp_g + p_g\Hat T_g$, which yields:
$$
\Hat T_g = - p_g^{-1} T_g p_g.
$$
If we write $p_g^{-t} := (p_g^t)^{-1} = (p_g^{-1})^t\Strut$ for
$g \in SO'_*(V)$, we then find that
$p_g + q_g \Hat T_g = p_g + q_g (q_g^t p_g^{-t})
 = (p_g p_g^t + q_g q_g^t) p_g^{-t} = p_g^{-t}\Strut$. So,
whenever $g^{-1},h,gh \in SO'_*(V)$, we have:
$$
\eqalignno{
T_{gh} &= q_{gh}p_{gh}^{-1} = (q_g + p_gT_h)(p_g + q_gT_h)^{-1}
 = (q_g + p_gT_h)(I - \Hat T_gT_h)^{-1} p_g^{-1}  \cr
&= q_g p_g^{-1} + (q_g + p_gT_h - q_g(I - \Hat T_gT_h))
                  (I - \Hat T_gT_h)^{-1} p_g^{-1} \cr
&= T_g + p_g^{-t} T_h (I - \Hat T_gT_h)^{-1} p_g^{-1}.
  & (2.3)  \cr}
$$

\smallskip

   A few words on traces and determinants are in order too.
If $A$ is a traceclass linear operator, we will denote its
{\it complex trace} $\sum_k \<e_k, A e_k>_{\scriptscriptstyle J}$
by $\Tr_\C A$; where $\{e_k\}_{k\in\N}$ is any orthonormal basis
on $(V,d,J)$. If $S,T$ are antilinear Hilbert--Schmidt operators
on~$V$, then $\Tr_\C(\comm ST)$ need not vanish; indeed,
$\Tr_\C(\comm ST) = \sum_k \<Se_k, Te_k> - \<Te_k, Se_k>$ is a
purely imaginary number. (The complex trace does, of course,
vanish on commutants of linear operators.) Likewise, we will use
$\detc$ to denote a {\it complex determinant\/}: if $A$ is a
linear operator on~$V$ and $V$ is finite-dimensional, we define
$\detc A$ to be the determinant of the matrix with entries
$\<e_i, Ae_j>$. When $V$ is infinite-dimensional, $\detc A$
still makes sense provided $A-I$ is traceclass, and the
following\cite{\Araki} may be adopted as a definition:
$\detc(\exp N) := \exp(\Tr_\C N)$ for traceclass~$N$.

\section III. Operators and ``Gaussians''
              on the fermion Fock space

\subsection III.1. Gaussian elements

We first recall briefly the construction of the fermion Fock
space. We fix a complex structure $J \in \J(V)$ and regard~$V$
as the complex Hilbert space $(V,d,J)$. Its {\it exterior
algebra\/} is $\La(V) := \bigoplus_{n=0}^\infty V^{\w n}$, where
$V^{\w n}$ is the complex vector space algebraically generated
by the alternating products $v_1 \w v_2 \w\cdots\w v_n$, with
$V^{\w 0} = \C$ by convention. The inner product on~$\La(V)$ is
given by $\braket{u_1 \w\cdots\w u_m}{v_1 \w\cdots\w v_n}
 := \delta_{mn} \det(\<u_k,v_l>)$.

   The vacuum $\Omega$ is a fixed unit vector in~$V^{\w 0}$. The
antisymmetric Fock space $\F_J(V)$ is the Hilbert-space
completion of~$\La(V)$; most of the time we shall write only
$\F(V)$ instead of $\F_J(V)$. An orthonormal basis is given by
the elements $\eps_K := e_{k_1} \w\cdots\w e_{k_r}$, where
$\{e_n\}$ denotes a fixed orthonormal basis for $(V,d,J)$, and
$K = \set{k_1,k_2,\dots,k_r}$ is any finite set of positive
integers written in increasing order: $k_1 < \cdots < k_r$ (with
$\eps_K = \Omega$ if $K$ is void).

   The fermion Fock space splits as $\F(V) = \F_0 \oplus \F_1$,
where $\F_0$ is the completion of the even part
$\bigoplus_{k=0}^\infty V^{\w(2k)}$ of the exterior algebra, and
$\F_1$ is the completion of
$\bigoplus_{k=0}^\infty V^{\w(2k+1)}$.

\smallskip

   If $T \in \Sk(V)$, the series
$$
H_T := \sum_{i,j=1}^\infty \<e_i,Te_j> e_i \w e_j
\eqno (3.1)
$$
converges in~$\F_0$ since
$\sum_{i,j=1}^\infty \abs{\<e_i,Te_j>}^2
 = \sum_{j=1}^\infty \<Te_j,Te_j> < \infty$, the sum being
independent of the basis chosen.

   We call {\it Gaussians\/} the following ``quadratic
exponential'' elements of $\F_0$:
$$
f_T := \exp^\w(\thalf H_T)
 := \sum_{m=0}^\infty {1\over 2^m m!} H_T^{\w m}.
\eqno (3.2)
$$
If $T_m \in \Sk(V)$ is determined by $T_m(e_{2m-1}) = - e_{2m}$,
$T_m(e_{2m}) = e_{2m-1}$, $T_m(e_j) = 0$ for other~$j$, then
$H_{T_m} = 2e_{2m-1} \w e_{2m}$, and
$f_{T_m} = \Omega + e_{2m-1} \w e_{2m}$. Furthermore, if
$T = T_1 + \cdots + T_m$, then
$H_T = 2 \sum_{k=1}^m e_{2m-1} \w e_{2m}$, and
$f_T = \Omega + e_1 \w\cdots\w e_{2m} + {}$(lower order terms).
It follows that the linear span of $\set{f_T : T \in \Sk(V)}$ is
dense in~$\F_0$.

\smallskip

   We can expand the Gaussian~$f_T$ in the orthonormal basis
$\{\eps_K\}$. First, note that
$$
{1\over 2^m m!} H_T^{\w m}
 = \sum_{\abs K=2m} {1\over 2^m m!} (\pm)_K \<e_{k_1},Te_{k_2}>
   \dots \<e_{k_{2m-1}},Te_{k_{2m}}> e_{k_1} \w\dots\w e_{k_{2m}}
\eqno (3.3)
$$
where $(\pm)_K$ is the sign of the permutation putting
$K = \{k_1,\dots,k_{2m}\}$ in increasing order, and $T_K$
denotes the skewsymmetric $2m \x 2m$ matrix with entries
$\<e_k,Te_{k'}>$ for $k,k'\in K$. Recall\cite{\Bourbaki} that
the {\it Pfaffian\/} of a skewsymmetric $2m\x 2m$ matrix~$A$ is
given by
$$
\Pf A := {1\over 2^m\,m!} \sum_{\sigma\in S_{2m}}
 \pm_\sigma\, a_{\sigma(1)\sigma(2)}\, a_{\sigma(3)\sigma(4)}
 \dots a_{\sigma(2m-1)\sigma(2m)},
$$
and satisfies the crucial property $(\Pf A)^2 = \det A$. If $A$
is a skewsymmetric $(2m+1) \x (2m+1)$ matrix, then $\det A = 0$,
so one defines $\Pf(A) := 0$. Thus we can rewrite~\eq{3.3} as
$$
{1\over 2^m m!} H_T^{\w m} = \sum_{\abs K=2m} \Pf(T_K) \eps_K.
$$
By convention, we take $\Pf(T_K) := 1$ when $K$ is the void set.
We thus arrive at the expansion for~\eq{3.2}:
$$
f_T = \sum_{K\,\rm finite} \Pf(T_K)\, \eps_K,
\eqno (3.4)
$$
where only the even subsets $K \subset \N$ contribute to the sum.

\smallskip

   It is shown in~Ref.~\Pressley\ that
$\sum_K \Pf(S_K) \Pf(T_K)$ is a square root of $\det(I - TS)$
whenever $S$, $T$ are skewsymmetric real matrices. For complex
matrices, the corresponding formula is:
$$
\biggl( \sum_K \Pf(S_K)^* \Pf(T_K) \biggr)^2 = \detc(I - TS).
$$
We may summarize the foregoing discussion as follows (we suppress
the subscript `$\C$' on fractional powers of complex determinants):

\proclaim Lemma.
Let $S,T \in \Sk(V)$ where $\dim V = 2n$ is finite. Then if
$\det^{1/2}(I - TS)$ de\-notes the square root of\/
$\detc(I - TS)$ which equals\/~$1$ when $S = 0$ or~$T = 0$, the
following expansion is valid:
$$
\det^{1/2}(I - TS) = \sum_K \Pf(S_K)^* \Pf(T_K),
\eqno (3.5)
$$
where the sum ranges over the principal submatrices of even
dimension.

\smallskip

   When $V$ is infinite dimensional and $T \in \Sk(V)$, we have
$$
\sum_{K\rm\,finite} \abs{\Pf(T_K)}^2 = \det^{1/2}(I - T^2)
 = \det^{1/2}(I + T^tT),
$$
using~\eq{3.5} for finite-rank $T$ and a limiting argument. The
series $\sum_{K\rm\,finite} \Pf(S_K)^* \Pf(T_K)$ converges for
$S,T \in \Sk(V)$ by the Schwarz inequality, so \eq{3.5} holds
for all $S,T \in \Sk(V)$.

   The inner product of two fermionic Gaussians follows at once
 from the expansions~\eq{3.4} and~\eq{3.5}:
$$
\<f_S,f_T> = \det^{1/2}(I - TS),  \sepword{for}  S,T \in \Sk(V).
\eqno (3.6)
$$

\subsection  III.2. Representing the field algebra

The basic object in quantization is the field algebra over the
space $(V,d)$, which is defined prior to any choice of the
complex structure. The field algebra over the space $(V,d)$ is
just the complexified Clifford algebra
$\Ag(V_\C) := \opname{Cl}(V,d) \otimes \C$, complete \wrt~the
natural (inductive limit) $C^*$-norm.\cite{\Emch} There is a
linear map $B\: V_\C \to \Ag(V_\C)$ ---the ``fermion field''---
satisfying $B(w^*) = B(w)^\7$ and
$$
\{B(v), B(v')\} = 2\,d(v,v')  \sepword{for all} v,v' \in V.
\eqno (3.7)
$$
Any $C^*$-algebra generated by a set of operators
$\set{B'(w) : w \in V_\C}$ satisfying the same rules is
isomorphic\cite{\Araki} to $\Ag(V_\C)$ via $B(w) \mapsto B'(w)$.

   One can construct a faithful irreducible representation
$\pi_J$ of $\Ag(V_\C)$ by the GNS cons\-truction with respect to
the ``Fock state'' $\omega_J$ determined by
$\omega_J(B(u)B(v)) := \<u,v>_{\scriptscriptstyle J}$;
see~Ref.~\Araki\ for details. It turns out that this is
equivalent to the standard representation of the canonical
anticommutation relations (CAR) on the fermion Fock space
$\F_J(V)$. The annihilation and creation operators for the free
fermion field may be defined as real-linear operators on~$\F(V)$:
$$
a_J(v) := \pi_JB(P_{-J}v),  \qquad   a_J^\7(v) := \pi_JB(P_Jv).
\eqno (3.8)
$$
Note that the vacuum sector is associated to the polarization:
$\pi_JB(w)\Omega = 0$ for $w\in W_0^*$. Since
$P_{\pm J} = \thalf(I \mp iJ)$, we thus have
$a_J(Jv) = -i\,a_J(v)$ and $a_J^\7(Jv) = i\,a_J^\7(v)$ for
$v \in V$; and $\pi_JB(v) = a_J(v) + a_J^\7(v)$. Since
$V^{\w 1} \subset \F(V)$ is the complex Hilbert space~$(V,d,J)$,
we have $iv = Jv$ in $V^{\w 1}$; thus
$a_J^\7(v)\Omega = \thalf v - \tihalf Jv = v$,
$a_J(v)\Omega = \thalf v + \tihalf Jv = 0$ in~$V^{\w 1}$. More
generally, $a_J^\7(v_1)a_J^\7(v_2) \dots a_J^\7(v_k)\Omega
 = v_1 \w v_2 \w\cdots\w v_k \in V^{\w k}$. It is
straightforward to check that
$\{\pi_JB(v), \pi_JB(v')\} = 2\,d(v,v')$ on vectors of the form
$v_1 \w\cdots\w v_k$. The canonical anticommutation relations
$\{a_J(v), a_J(v')\} = 0, \{a_J(v), a_J^\7(v')\} = \<v,v'>$
follow directly from~\eq{3.8}.

   In summary, to each complex structure, in principle there
corresponds a different identification of the Clifford algebra
with a CAR algebra. Since $\pi_J$ is faithful, as long as we
consider a single complex structure ---thus a single Fock space
representation--- we will no longer need to distinguish between
$B(v)$ and $\pi_JB(v)$ in the notation. This amounts to
regarding $\Ag(V_\C)$ as the algebra of field operators
on~$\F_J(V)$.

   If $U \in U_J(V)$, one defines the unitary operator $\Ga_J(U)$
on~$\F_J(V)$ by the usual rule $\Ga_J(U) := U\w U\w\cdots\w U$
with $\Ga_J(U)\Omega := \Omega$. Using
$B(v) = a_J(v) + a_J^\7(v)$, we get the intertwining property:
$$
\Ga_J(U) B(v) \Ga_J(U)^{-1} = B(Uv).
\eqno (3.9)
$$

   Observables of the one-particle theory are given by elements
of the Lie algebra $\ol'(V)$ of the restricted orthogonal group
consisting of skewsymmetric real-linear operators
$X \in \Endr(V)$. We can quantize at the present stage
the Lie algebra elements commuting with $J$. If $X\in\ol'(V)$
and $\comm XJ = 0$, then it is immediately seen that $-JX$ is
selfadjoint on~$(V,d,J)$. If $A$ is a selfadjoint operator
on~$(V,d,J)$, then the quantized operator or {\it current\/}
$d\Ga_J(A)$ on $F(V)$ defined as
$d\Ga_J(A) := {d\over dt}\bigr\vert_{t=0} \Ga_J(\exp(itA))$ is
explicitly given by
$$
d\Ga_J(A)(v_1 \w v_2 \w\cdots\w v_n)
 := \sum_{k=1}^n v_1 \w\cdots\w v_{k-1} \w Av_k \w v_{k+1}
     \w\cdots\w v_n,
$$
whenever $v_1,v_2,\dots,v_n \in \Dom(A)$, and
$d\Ga_J(A)\Omega := 0$. Such vectors span a dense subspace of
$\F(V)$ which is invariant under the one-parameter group
$\Ga_J(\exp(itA))$, and hence is a core for
$d\Ga_J(A)$.\cite{\Taylor} In the terminology of~Ref.~\Pressley,
$\Ga$ is a {\it positive-energy\/} representation of~$U_J(V)$.
We remark that $(\F_J(V), \pi_JB,\Omega,\Ga)$ is a {\it full
quantization\/} of~$(V,d)$ in the sense of
Segal.\cite{\Segalcwave} That is to say: (a)~$\F_J(V)$ is a
separable Hilbert space; (b)~$\pi_JB(V)$ is a system of
selfadjoint operators on this Hilbert space,
satisfying~\eq{3.7}; (c)~$\Omega$ is a unit vector in~$\F_J(V)$
which is a cyclic vector for~$\pi_JB(V)$; and (d)~$\Ga$ is a
unitary representation of~$U_J(V)$ intertwining $\pi_JB(V)$, for
which $\Omega$ is stationary, such that $d\Ga(A)$ is positive
on~$\F_J(V)$ whenever $A$ is positive on the Hilbert space
$(V,d,J)$.

\section IV. The pin and spin representations

\subsection IV.1. Vacuum sectors and the
                  Shale--Stinespring criterion

Given such a full quantization of~$(V,d)$, then for any
$g \in O(V)$ it follows from~\eq{3.7} that $v\mapsto \pi_JB(gv)$
defines another full quantization acting on the same Hilbert
space. Indeed, $w \mapsto B(gw)$ (for $w \in V_\C$) extends to a
$*$-automorphism of the CAR algebra~$\Ag(V_\C)$. We then ask
whether these two quantizations are unitarily equivalent, i.e.,
whether this \hbox{$*$-automorphism} is unitarily implementable
on~$\F_J(V)$. For a given $g\in O(V)$, we seek a unitary
operator $\mu(g)$ on~$\F(V)$ so that, extending~\eq{3.9}:
$$
\mu(g) B(v) = B(gv) \mu(g),  \sepword{for all} v \in V.
\eqno (4.1)
$$

   The complex structure $J$ is transformed to $gJg^{-1}$; the
creation and annihiliation operators undergo a {\it Bogoliubov
transformation\/}:
$$
a_{gJg^{-1}}(gv) = a_J(p_gv) + a_J^\7(q_gv),  \qquad
a_{gJg^{-1}}^\7(gv) = a_J(q_gv) + a_J^\7(p_gv).
\eqno (4.2)
$$
Were $\mu(g)$ to exist, we would then have
$$
\mu(g) a_J(v) = a_{gJg^{-1}}(gv) \mu(g),  \qquad
\mu(g) a_J^\7(v) = a_{gJg^{-1}}^\7(gv) \mu(g).
$$

   Thus the {\it out-vacuum\/} $\mu(g)\Omega$ is annihilated
by~$a_{gJg^{-1}}(gv)$, for all $v \in V$. Since
$\set{a_J(v) : v\in V}$ vanishes only on scalar multiples
of~$\Omega$, the kernel of $\set{a_{gJg^{-1}}(gv) : v \in V}$ is
one-dimensional also. Indeed, since
$\set{a_{gJg^{-1}}(gv) : v \in V} = \set{B(z) : z \in gW_0^*}$,
this kernel, if it is nonzero, must be the vacuum sector for the
polarization~$gW_0^*$.

   We first of all observe that when $g \in SO'_*(V)$, this
vacuum sector is generated by a Gaussian element:
$$
a_{gJg^{-1}}(gv) f_{T_g} = 0  \sepword{for all} v \in V
  \sepword{if}  g \in SO'_*(V).
$$
Indeed, if we replace $v$ by~$p_g^{-1}v$ and use~\eq{4.2}, we
need only show $(a_J(v) + a_J^\7(T_gv)) f_{T_g} = 0$ for all
$v \in V$; we can moreover take $v = e_1$, the first element in
an orthonormal basis for~$V$. From~\eq{3.4}, it is not hard to
establish that
$$
a_J(e_1) f_T + a_J^\7(Te_1) f_T = 0
  \sepword{for any}  T \in \Sk(V),
\eqno (4.3)
$$
on expanding the Pfaffians with a little linear algebra.

   From this we see that $a_{gJg^{-1}}(gv) \Psi = 0$ for all~$v$
implies that $\Psi = c_g f_{T_g}$, for some constant $c_g$.

\smallskip

   If $g \in O'(V)$ with
$\dim(\ker p_g) = \dim(\ker p_g^t) = n > 0$, one proceeds as
follows.\cite{\Ruijsenaars} Let $\{e_1,\dots,e_n\}$ be an
orthonormal basis for the subspace $\ker p_g^t$. Let $r_k$ be the
reflection of~$(V,d)$ for which
$r_k(e_k + Je_k) = - (e_k + Je_k)$, $r_k(v) = v$ if
$d(v, e_k + Je_k) = 0$, and let $r = r_1\dots r_n$. Then
$rg \in SO'_*(V)$, and one can check that
$$
(a_J(p_gv) + a_J^\7(q_gv))
 \lbrack e_1 \w\cdots\w e_n \w f_{T_{rg}} \rbrack = 0.
$$
Using~\eq{3.4} and recalling that $\Pf(T_K) = 0$ unless $K$ is
even, we see that out-vacuum sector
$\C(e_1 \w\cdots\w e_n \w f_{T_{rg}})$ lies in $\F_0(V)$ or
$\F_1(V)$ according as $n$ is even or~odd.

   We turn now to the necessary conditions for the existence of
the unitary operators $\mu(g)$. Since the representation $\pi_J$
of the CAR algebra is irreducible, \eq{4.1} shows that $\mu(g)$
is determined by $\mu(g)\Omega$; and this vector lies in the
one-dimensional subspace spanned by
$e_1 \w\cdots\w e_n \w f_{T_{rg}}$ if $g \in O'(V)$. Shale and
Stinespring\cite{\ShaleStine} have shown that a necessary
condition for a vacuum vector for $B(gW_0^*)$ to exist inside
$\F_J(V)$ is that $\comm Jg$ be Hilbert--Schmidt. The proof of
the Shale--Stinespring theorem in the present treatment amounts
to showing that such a vector must necessarily be a multiple of
$e_1 \w\cdots\w e_n \w f_{T_{rg}}$. In order that this expression
be convergent, $n$~must be finite and $f_{T_{rg}}$ must lie
in~$\F(V)$; by~\eq{3.6}, the operator $I - T_{rg}^2$ must have a
determinant, i.e., $T_{rg}$ must be Hilbert--Schmidt. Both
conditions are fulfilled iff $g\in O'(V)$, since the finiteness
of~$n$ amounts to $p_g$ being Fredholm. From now on, we will
assume that $g\in O'(V)$.

   Thus $\mu(g)\Omega = c_g e_1 \w\cdots\w e_n \w f_{T_{rg}}$.
Unitarity demands that
$$
1 = \<\mu(g)\Omega,\mu(g)\Omega>
  = \abs{c_g}^2 \<f_{T_{rg}}, f_{T_{rg}}>
  = \abs{c_g}^2 \det^{1/2}(I - T_{rg}^2).
$$
Thus $\abs{c_g} = \det^{-1/4}(I - T_{rg}^2)$.  We now {\it fix
the phase\/} by taking $c_g > 0$, i.e.,
$$
c_g :=  \det^{-1/4}(I - T_{rg}^2).
$$
(Although there is some arbitrariness in the choice of~$r$, it
happens that $T_{rg}$ vanishes on $\ker p_g^t$ and one finds that
$c_g = \det^{1/4}(p_g^t p_g\bigr\vert_{(\ker p_g)^\perp})$,
independent of~$r$.)

\subsection IV.2. Construction of the pin
                  and spin representations

We are now ready to write down the pin representation
of~$O'(V)$, along the lines of Pressley and
Segal.\cite{\Pressley} Their treatment is explicit only for a
finite-dimensional~$V$, and can be checked to coincide with the
more usual Clifford-algebra construction of the pin
repre\-sen\-ta\-tion\cite{\Lawson,\Sirius} for the double cover
$\opname{Pin}(2m)$ of~$O(2m)$. The technique is to permute
Gaussian elements, paying due regard to exceptional cases. We
present it here in a somewhat different form from~Ref.~\Pressley,
in order to make clear that the infinite-dimensional case poses
no extra difficulty. An excellent treatment from the
Clifford-algebra viewpoint was given by Carey and
Palmer.\cite{\CareyPalmer,\Palmer} For supersymmetric
generalizations, consult the sketch in~Ref.~{\Popov} and the
detailed treatment of a finite dimensional case
in~Ref.~{\Borthwick}.

   The first step is to note that there is a local action of
$SO'(V)$ on $\Sk(V)$, given by:
$$
g\.S := (q_g + p_gS)(p_g + q_gS)^{-1}.
$$
(If $p_g$ is invertible, so is $p_g + q_gS$ for $S$ small enough
in the Hilbert--Schmidt norm.) Since the relations~\eq{2.1} yield
$$
(p_g + q_gS)^t (q_g + p_gS) = - (q_g + p_gS)^t (p_g + q_gS)
$$
whenever $S^t = - S$, it follows that $g\.S$ is also skew, and it
is clearly antilinear and Hilbert--Schmidt; thus
$g\.S \in \Sk(V)$. It is readily checked that $gh\.S = g\.(h\.S)$
whenever $h\.S$ and $gh\.S$ are defined; the group action is
``local'' since they may be undefined in particular cases.
 From~\eq{2.3} with $T_h = S$, we obtain the useful alternative
form:
$$
g\.S = T_g + p_g^{-t} S (I - \Hat T_gS)^{-1} p_g^{-1}.
\eqno (4.4)
$$

   For a given $g \in SO'_*(V)$, other than the identity, those
$S$ for which $p_g + q_gS$ is invertible form an open
neighbourhood of zero in~$\Sk(V)$, so
$\set{f_S : g\.S \rm\ exists}$ spans a dense subset of~$\F_0(V)$.
Thus we may define a unitary operator on $\F_0(V)$ by the
prescription:
$$
\mu(g) f_S := c_g \phi_g(S) f_{g\.S},
\eqno (4.5)
$$
where $\phi_g(S) \in \C$ is to be chosen to make $\mu(g)$
unitary. A suitable choice is:
$$
\phi_g(S) := \det^{1/2}(I - S\Hat T_g).
\eqno (4.6)
$$
We have $\phi_g(0) = 1$ and
$c_g^2 \<\phi_g(S) f_{g\.S}, \phi_g(T) f_{g\.T}> = \<f_S,f_T>$,
using~\eq{3.6}.

   From the definition~\eq{4.6} and the formula~\eq{4.4}, it
follows that
$$
\eqalignno{
\phi_{gh}(S)
&= \det^{1/2}(I - S(\Hat T_h + p_h^{-1}
                    \Hat T_g (I-T_h\Hat T_g)^{-1} p_h^{-t}))\cr
&= \det^{1/2}(p_h^{-t}(I - S\Hat T_h)p_h^t - p_h^{-t} S p_h^{-1}
                    \Hat T_g (I - T_h\Hat T_g)^{-1}))  \cr
&= \det^{-1/2}(I - T_h \Hat T_g)\,
   \det^{1/2}((I - S\Hat T_h) p_h^t (I - T_h\Hat T_g) p_h^{-t}
               - S p_h^{-1} \Hat T_g p_h^{-t})  \cr
&= \det^{-1/2}(I - T_h \Hat T_g)\, \phi_h(S)\,
   \det^{1/2}(I - T_h\Hat T_g - p_h^{-t}
                    (I-S\Hat T_h)^{-1} S p_h^{-1} \Hat T_g) \cr
&= \det^{-1/2}(I - T_h\Hat T_g)\,\phi_h(S)\, \phi_g(h\.S),  \cr}
$$
whenever $g,h,gh \in SO'_*(V)$, provided $h\.S$ and $gh\.S$ both
exist. The set of such $S$ is a neighbourhood of zero in
$\Sk(V)$, so that the corresponding Gaussians $f_S$ are total in
$\F_0(V)$; we thus arrive at
$$
\mu(g) \mu(h) = c(g,h) \mu(gh),
\eqno (4.7)
$$
where the {\it cocycle\/} $c(g,h)$ is given by
$$
c(g,h) := c_g c_h c_{gh}^{-1} \det^{1/2}(I - T_h\Hat T_g)
 = \exp(i\arg\det^{1/2}(I - T_h\Hat T_g)).
\eqno (4.8)
$$

   Thus the $g \mapsto \mu(g)$ is a {\it projective\/}
representation of the restricted orthogonal group. Actually, its
definition is incomplete, since in~\eq{4.5} we have defined
$\mu(g)$ only for $g \in SO'_*(V)$. To finish the job, we must
give the formulae for $\mu(g) \Psi$ and the respective cocycles,
when $g = rh$ with $h \in SO'_*(V)$, $r$ is a product of
reflections in~$O'(V)$, and $\Psi \in \F_0(V)$ or $\F_1(V)$. This
involves a considerable amount of bookkeeping; we will summarize
the results.

\smallskip

   If $v \in V$ is a unit vector, i.e., $d(v,v) = 1$, we define
$r_v \in O'(V)$ by
$$
r_v(u) := 2 d(v,u) v - u.
\eqno (4.9)
$$
Notice that $-r_v$ is the reflection across the hyperplane
orthogonal to~$v$. It is an improper orthogonal transformation,
since $\ker p_{r_v}$ is one-dimensional. We define simply:
$$
\mu(r_v) := B(v).
$$
If $u,v$ are unit vectors in~$V$ with $\<u,v> \neq 0$, then
$s = r_u r_v \in SO'_*(V)$. Then
$$
\mu(r_u) \mu(r_v) \Omega = c(r_u,r_v) \mu(r_u r_v) \Omega
 = \<u,v> \Omega + u \w v,
$$
with
$$
c(r_u,r_v) = \exp(i\arg\<u,v>).
$$
One then checks that
$\mu(r_u) \mu(r_v) \Psi = c(r_u,r_v) \mu(r_u r_v) \Psi$ for all
$\Psi \in \F_0(V)$.

   We complete now the definition of $\mu(g)$ for
$g \in SO'_*(V)$ by defining it on~$\F_1(V)$. In order to achieve
$\mu(g) B(v) = B(gv) \mu(g)$ on~$\F_0(V)$, we must set
$$
\mu(g) (B(v) f_S) := B(gv) \mu(g) f_S,
\eqno (4.10)
$$
for $S \in \Sk(V)$ such that $g\.S$ is defined. We are also free
to define
$$
\mu(gr_u) := \mu(g) \mu(r_u)  \sepword{for}  g \in SO'_*(V),
$$
and consequently $c(g,r_u) := 1$, so that \eq{4.10} yields
$\mu(gr_u) f_S = B(gu) \mu(g) f_S$. It can be verified that these
partial definitions are consistent. Now from~\eq{3.7} we obtain
$B(u) B(v) = B(r_uv) B(u)$, and one must then check that
$\mu(gr_u) B(v) = B(gr_uv) \mu(gr_u)$; thus \eq{4.1} holds for
elements of the form $g = hr_u$ with $h \in SO'_*(V)$. Since
$hr_uh^{-1} = r_{hu}$, we can equivalently say that it holds for
elements of the form $r_vh$, $h\in SO'_*(V)$.

   The general case now follows easily. We can always write
$g = rh$ where $p_h$ is invertible and $r = r_{e_1}\dots r_{e_n}$
is a product of $n$~elements of the form~\eq{4.9}, with
$\{e_1,\dots,e_n\}$ an orthonormal basis of $\ker p_g^t$; we then
define
$$
\mu(g) = B(e_1) \dots B(e_n) \mu(h).
$$
In particular, $\mu(g)\Omega = e_1 \w\cdots\w e_n \w f_{T_h}$, as
expected.

   The cocycle~\eq{4.7} is extended to all of $O'(V)$ as follows.
We define $c(g,r_u) := 1$ if $g \in SO'_*(V)$ and $r_u$ is of the
form~\eq{4.9}. We set $c(gr_u,r_v) := c(g, r_u r_v) c(r_u,r_v)$
if $d(u,v) \neq 0$; otherwise we are free to take $c(g,r_u r_v)$,
$c(r_u,r_v)$, $c(gr_u, r_v)$ all equal to~1. In general, we take
$\mu(g,r_{e_1}\dots r_{e_k}) = 1$ if $g \in SO'(V)$ and the $e_i$
are an orthonormal set of vectors in~$V$. The cocycle equation
$c(g,h) c(gh,k) = c(g,hk) c(h,k)$ then determines the remaining
values of~$c(g,h)$, in such a way that~\eq{4.7} remains valid for
all $g,h \in O'(V)$.

   We have now obtained the full ``pin representation''
of~$O'(V)$ on~$\F(V)$. By cons\-truc\-tion, it is an irreducible
projective representation. Its restriction to~$SO'(V)$ has two
orthogonal irreducible subspaces, namely $\F_0(V)$ and~$\F_1(V)$;
this is the ``spin representation'' of $SO'(V)$.

\section  V. Fermionic anomalies

\subsection V.1. The infinitesimal spin representation and the
                 quantization procedure

The spin representation allows us to quantize all elements of the
Lie algebra of the restricted orthogonal group $O'(V)$. If
$X\in\ol'(V)$ we write $C_X := \thalf(X - JXJ)$,
$A_X := \thalf(X + JXJ)$ for the linear and antilinear parts
of~$X$. If $t \mapsto \exp tX$ is a one-parameter group with
values in~$SO'(V)$, then $p_{\exp tX}$ is invertible for small
enough~$t$, and
$$
\ddto t p_{\exp tX} = C_X,  \qquad  \ddto t T_{\exp tX} = A_X.
$$
In particular, $A_X$ is Hilbert--Schmidt. Thus the antilinear
part of~$\ol'(v)$ is just $\Sk(V)$.

   We define the {\it infinitesimal spin representation\/}
$\dm(X)$ of $X\in \ol'(V)$ by:
$$
\dm(X)\Psi := \ddto t e^{i\theta_X(t)} \mu(\exp tX) \Psi
\eqno (5.1)
$$
for $\Psi \in \F(V)$, where $\theta_X(t)$ is such that
$t \mapsto e^{i\theta_X(t)} \mu(\exp tX)$ is a homomorphism.

   Writing $g(t) := \exp tX$ for small~$t$, we obtain
$$
\dm(X) \Omega
 = \ddto t e^{i\theta_X(t)} c_{g(t)} f_{T_{g(t)}}
 = i\theta'_X(0) \Omega + H_{A_X}.
$$

   Recall that if $A_X = 0$, the quantized counterpart of $X$ is
given by $d\Gamma(-JX)$, for which $d\Gamma(-JX)\Omega = 0$. The
vacuum expectation value of $-i\dm(X)$ is
$-i \<\Omega, \dm(X)\Omega> = \theta'_X(0)$. Since we may choose
$\theta'_X(0)$ arbitrarily, we set $\theta'_X(0) = 0$ for
all~$X \in \ol'(V)$. Hence the quantization rule
$X \mapsto -i\dm(X)$ is uniquely specified by~\eq{5.1} together
with the condition of vanishing vacuum expectation values.

   We shall write $dG(X) := -i \dm(X)$, for $X \in \ol'(V)$, to
denote our quantization rule. As we have just remarked, one has
$dG(X) = d\Ga(-JX)$ whenever the latter makes sense: this is
already clear from the fact that the spin representation
generalizes the intertwining property~\eq{3.9}.

   To be more precise about the domains of the (unbounded)
operators $\dm(X)$ on~$\F(V)$, we recall that for
any~$S \in \Sk(V)$, both $p_{\exp tX}$ and
$(p_{\exp tX} + q_{\exp tX}S)$ will be invertible for small~$t$,
and so we have:
$$
\eqalignno{
\dm(X) f_S
&= \ddto t e^{i\theta_X(t)} c_{g(t)}\phi_{g(t)}(S) f_{g(t)\.S} \cr
&= \ddto t \phi_{g(t)}(S) f_S
   + \sum_K \ddto t \Pf((g(t)\.S)_K)\, \eps_K \cr
&= \thalf \Tr_\C\lbrack SA_X\rbrack\, f_S
   + \thalf H_{\psi(X,S)} \w f_S,
& (5.2)  \cr}
$$
where $\psi(X,S) = \comm{C_X}S + A_X - SA_XS$. This shows that
$f_S \in \Dom(\dm(X))$ ---and in fact a similar calculation shows
that the application $X \mapsto \dm(X)f_S$ is differentiable.

   In like manner, the vectors $B(v)f_S$ lie in the domain
of~$\dm(X)$; indeed,
$$
\eqalignno{
\dm(X) (B(v)f_S)
&= \ddto t e^{i\theta_X(t)}
     c_{g(t)} \phi_{g(t)}(S) B(g(t)v) f_{g(t)\.S} \cr
&= B(v) \dm(X) f_S + \ddto t B((\exp tX)v) f_S  \cr
&= B(v) \dm(X) f_S + B(Xv) f_S
& (5.3) \cr}
$$
by real-linearity of~$B$. In the same way we get
$$
\dm(X) B(v) (B(u)f_S) = B(v) \dm(X) (B(u)f_S) + B(Xv) (B(u)f_S).
\eqno (5.4)
$$
Combining \eq{5.3} and~\eq{5.4}, we have the fundamental
commutation relations:
$$
\comm{\dm(X)}{B(v)} = B(Xv)
\eqno (5.5)
$$
as an operator-valued equation valid on the dense domain
in~$\F(V)$ generated by all~$f_S$ and~$B(u)f_S$. Indeed, \eq{5.5}
is the smeared expression for the formal commutation relations
between field operators and currents; this justifies the name
``currents'' for the quantized observables.

\subsection V.2. Schwinger terms and cyclic cohomology

The extended orthogonal group $\Onda{O'}(V)$ is the
one-dimensional central extension of $O'(V)$ by~$U(1)$ whose
elements are pairs $(g,\la)$, where $g \in O'(V)$,
$\la \in U(1)$, with group law
$$
(g,\la)\.(h,\la') = (gh, \la\la' c(g,h)),
$$
so that $(g,\la) \mapsto \la\mu(g)$ is a {\it linear\/} unitary
representation of the extended group. Its Lie algebra
$\Onda{\ol'}(V)$ is a one-dimensional central extension
of~$\ol'(V)$ by~$i\R$, with commutator
$$
\comm{(X,ir)}{(Y,is)} := (\comm XY, \a(X,Y)),
$$
where
$$
\a(X,Y) = {d^2\over dt\,ds}\biggr\vert_{t=s=0} c(\exp sX,\exp tY)
 - {d^2\over dt\,ds} \biggr\vert_{t=s=0} c(\exp tY, \exp sX).
\eqno (5.6)
$$
The Lie algebra cocycle $\a$ has the physical meaning of an
anomalous commutator or {\it Schwinger term}. Indeed, if
$X,Y \in \ol'(V)$, the Campbell--Baker--Hausdorff formula gives:
$$
\a(X,Y) = \comm{\dm(X)}{\dm(Y)} - \dm(\comm XY).
$$

\proclaim Proposition.
If $X,Y \in \ol'(V)$, then
$$
\a(X,Y) = -\thalf \Tr_\C( \comm{A_X}{A_Y} ).
\eqno (5.7)
$$

\Proof:
The linear and antilinear parts of
$\comm XY = \comm{C_X + A_X}{C_Y + A_Y}$ are given by
$C_{\comm XY} = \comm{C_X}{C_Y} + \comm{A_X}{A_Y}$,
$A_{\comm XY} = \comm{A_X}{C_Y} + \comm{C_X}{A_Y}$. The commutator
$\comm{\dm(X)}{\dm(Y)}$ may be computed from the quantization
formula~\eq{5.9} by substituting \eq{4.5}; using the CAR, one
finds that
$$
\eqalignno{
\comm{a^\7 A_X a^\7}{a^\7 C_Y a} &= a^\7 \comm{A_X}{C_Y} a^\7, \cr
\comm{a^\7 C_X a}{a A_Y a} &= a \comm{C_X}{A_Y} a,  \cr
\comm{a^\7 C_X a}{a^\7 C_Y a} &= a^\7 \comm{C_X}{C_Y} a,  \cr
\comm{a^\7 A_X a^\7}{a A_Y a} + \comm{a A_X a}{a^\7 A_Y a^\7}
&= -4\, a^\7 \comm{A_X}{A_Y} a + 2 \Tr_\C( \comm{A_X}{A_Y} ), \cr}
$$
It then follows that $\comm{\dm(X)}{\dm(Y)}
 = \dm(\comm XY) - \thalf \Tr_\C(\comm{A_X}{A_Y})$.   \qed

\smallskip

One can obtain the same result directly from~\eq{5.6}.

\smallskip

   The formula~\eq{5.7} yields the Schwinger term directly from
the obstruction to line\-arity of the pin representation. When
$V$ is finite-dimensional, the following reformulation is
possible: since the linear commutant $\comm{C_X}{C_Y}$ is
traceless, \eq{5.7} reduces to
$\a(X,Y) = -\thalf \Tr_\C\lbrack C_{\comm XY}\rbrack$, which is a
trivial cocycle; thus the pin representation appears as a
{\it linear\/} representation of a double covering group
$\opname{Pin}(2m)$ of~$O(2m)$. In the infinite-dimensional case,
this is no longer true, since $\comm{C_X}{C_Y}$ is in general not
traceclass.

   The Lie algebra cocycle $\a$ turns out to be also a cocycle
for the {\it cyclic cohomology\/} of
Connes\cite{\Connes,\Sirius}; this has been pointed out by
Araki.\cite{\Araki} We start from the observation that
$$
\a(X,Y) = {i\over4} \Tr(J\comm JX\comm JY)
  \sepword{for} X,Y \in \ol'(V).
$$
Here $\Tr$ is the usual trace over the real Hilbert space $(V,d)$.
Notice that $\Tr(J\comm JY\comm JX) = \Tr(\comm JX J\comm JY)
 = - \Tr(J\comm JX\comm JY)$; antisymmetrization of the right
hand side yields ${i\over2}\Tr(J\comm{A_X}{A_Y})
 = -\thalf \Tr_\C(\comm{A_X}{A_Y}) = \a(X,Y)$. Now a cyclic
1-cochain is simply an antisymmetric bilinear form~$\omega$; and
the cyclic coboundary operator $b$, defined by
$$
b\omega(X,Y,Z) := \omega(XY,Z) - \omega(X,YZ) + \omega(ZX,Y),
$$
yields $b\a(X,Y,Z)
 = (i/4) \sum_{\rm cyclic} \Tr(J \comm J{XY} \comm JZ) = 0$, so
the fermionic Schwinger term $\a$ is a cyclic 1-cocycle.
Higher-order cyclic cocycles also appear, somewhat mysteriously,
in current algebras,\cite{\GrosseMad} in a manner closely related
to the present approach.\cite{\Mickelssonbis}

\subsection V.3. Anomalies

   The group $O'(V)$ acts on $\Onda{\ol'}(V)$ by the adjoint
action of the central extension:
$$
\Onda\Ad(g) : (X,ir) \mapsto (\Ad(g)X, ir + \ga(g,X)),
$$
where the {\it anomaly\/} $\ga(g,X) \in i\R$ depends linearly
on~$X$. Since $\Onda\Ad(g) \comm{(X,ir)}{(Y,is)}
 = \comm{\Onda\Ad(g)(X,ir)}{\Onda\Ad(g)(Y,is)}$, we obtain
$$
\ga(g, \comm XY) = \a(\Ad(g)X, \Ad(g)Y) - \a(X,Y),
$$
for $X,Y \in \ol'(V)$. Therefore,
{\it the anomaly is determined by the Schwinger terms}. Moreover,
for $g \in O'(V)$, $X \in \ol'(V)$, we have:
$$
\ga(g,X) = \mu(g) \dm(X) \mu(g)^{-1} - \dm(\Ad(g)X).
\eqno (5.8)
$$

   Indeed, from~\eq{5.1} we obtain
$$
\eqalignno{
\mu(g) \dm(X) \mu(g)^{-1}
&= \ddto t e^{i\theta_X(t)} c(g, \exp tX) c(g\exp tX, g^{-1})
     \mu(g \exp tX g^{-1})   \cr
&= \ddto t c(g, \exp tX) c(g\exp tX, g^{-1}) + \dm(\Ad(g)X)
& (5.9)  \cr}
$$
(where we have used
$\dot\theta_X(0) = \dot\theta_{\Ad(g)X}(0) = 0$). Thus the right
hand side of~\eq{5.8} is an (imaginary) scalar; call it
$\ga'(g,X)$. That $\ga'(g,\comm XY) = \ga(g,\comm XY)$ in general
is clear from:
$$
\eqalign{
\ga'(g,\comm XY)
&= \mu(g) \dm(\comm XY) \mu(g)^{-1}
   - \dm(\comm{\Ad(g)X}{\Ad(g)Y})  \cr
&= \mu(g) \comm{\dm(X)}{\dm(Y)} \mu(g)^{-1} - \a(X,Y)  \cr
&\qquad - \comm{\dm(\Ad(g)X)}{\dm(\Ad(g)Y)}
   + \a(\Ad(g)X, \Ad(g)Y)  \cr
&= \comm{\dm(\Ad(g)X) + \ga'(g,X)}{\dm(\Ad(g)Y) + \ga'(g,Y)} \cr
&\qquad- \comm{\dm(\Ad(g)X)}{\dm(\Ad(g)Y)} + \ga(g,\comm XY). \cr}
$$

   It is now straightforward to compute the fermionic anomaly.

\proclaim Proposition.
For $g \in SO'_*(V)$, $X \in \ol'(V)$, we have
$$
\ga(g,X) = - \thalf \Tr_\C \bigl( (I - \Hat T_g^2)^{-1}
  (\comm{A_X}{\Hat T_g} - \Hat T_g \comm{C_X}{\Hat T_g}) \bigr).
\eqno (5.10)
$$

\Proof:
Let us write $h := \exp tX$. From~\eq{5.9}, we see that
$\ga(g,X)$ is given by the formula
$\ga(g,X) = {d\over{dt}}\bigr\vert_{t=0} c(g,h) c(gh, g^{-1})$.
The right hand side equals
$$
\eqalignno{
&\ddto t \exp( i\arg(\det^{1/2}(I - T_h \Hat T_g)
                   + \det^{1/2}(I - \Hat T_g \Hat T_{gh})))  \cr
&= i \Im \biggl( \ddto t \det^{1/2}(I - T_h \Hat T_g)
    + \det^{-1/2}(I - \Hat T_g^2)
      \ddto t \det^{1/2}(I - \Hat T_g\Hat T_{gh}) \biggr)  \cr
&= \tihalf \Im \Tr_\C \biggl( \ddto t (I - T_h \Hat T_g)
    + (I - \Hat T_g^2)^{-1}
      \ddto t (I - \Hat T_g\Hat T_{gh}) \biggr)  \cr
&= - \tihalf \Im \Tr_\C \biggl( A_X \Hat T_g
    + (I - \Hat T_g^2)^{-1} \Hat T_g
      \ddto t \Hat T_{gh} \biggr)  \cr
&= - \tihalf \Im \Tr_\C \biggl( A_X \Hat T_g
    + (I - \Hat T_g^2)^{-1} \Hat T_g
      \ddto t (\Hat T_h + p_h^{-1}\Hat T_g
               (I-T_h\Hat T_g)^{-1} p_h^{-t}) \biggr) \cr
&= - \thalf \Tr_\C \biggl( (I - \Hat T_g^2)^{-1}
      (\comm{A_X}{\Hat T_g} - \Hat T_g \comm{C_X}{\Hat T_g} )
      \biggr).  \cr}
$$
Notice that the linear operator
$\comm{A_X}{\Hat T_g} - \Hat T_g \comm{C_X}{\Hat T_g}
 = \comm{A_X - \Hat T_g C_X}{\Hat T_g}$, as a commutator of two
antilinear operators, has purely imaginary trace.   \qed

\smallskip

   Formula~\eq{5.10} seems to be new: appraisal of its strengths
and weaknesses is in order. Within its range of validity it is
completely general. This makes it very useful in conformal field
theory. For instance, the anomalous conservation laws for the
energy-momentum tensor and other observables, obtained in
Ref.~{\Maderner} by a direct procedure, can be computed with less
labour from~\eq{5.10} by the well known trick of embedding the
Virasoro group in an infinite dimensional orthogonal group. The
appearance of the {\it commuting part\/} of~$X$ in~\eq{5.10} also
deserves a comment: whereas observables that are linear (in the
sense of commuting with the complex structure) have nonanomalous
commutators for their corresponding quantum currents, they still
suffer in general from anomalous transformation laws. On the
other hand, \eq{5.10} is not applicable to the anomalies of the
charge and chiral charge conservation laws, which are partly
``topological'' in nature: see our treatment in subsection~VII.3.

\section VI. The spin representation in terms of field operators

\subsection VI.1. Currents

We reexpress the quantization prescription of subsection~V.1 in
the more congenial Fock space language. Given orthonormal bases
$\{e_j\}$, $\{f_k\}$ of $(V,d,J)$, let us introduce the
quadratic expressions:
$$
\eqalignno{
a^\7 T a^\7
&:= \sum_{j,k} a_J^\7(f_k) \<f_k,Te_j> a_J^\7(e_j),  \cr
a T a &:= \sum_{j,k} a_J(e_j) \<Te_j,f_k> a_J(f_k),
& (6.1)  \cr
a^\7 C a &:= \sum_{j,k} a_J^\7(f_k) \<f_k,Ce_j> a_J(e_j).
\cr}
$$
Formulae~\eq{6.1} are independent of the orthonormal bases used
iff $T$ is antilinear and skew and $C$ is linear as
ope\-ra\-tors in~$\Endr(V)$. We also have
$$
a^\7 T a^\7 \,\Omega = \sum_{j,k} \<e_k,Te_j>\, e_k \w e_j = H_T,
\eqno (6.2)
$$
which lies in~$\F_0(V)$ if and only if $T$ is Hilbert--Schmidt;
and more generally, $a^\7 T a^\7 \Psi = H_T \w \Psi$ for
$\Psi \in \F(V)$. Thus the series for $a^\7 T a^\7$ is meaningful
and defines a bounded operator on~$\F(V)$ iff $T \in \Sk(V)$.

   One easily sees that $a T a$ is the adjoint of~$a^\7 T a^\7$.
If $T,S \in \Sk(V)$, then we find that:
$$
(a T a) f_S = H_{STS} \w f_S - \Tr_\C \lbrack ST\rbrack\, f_S.
$$
If $C$ is a skewadjoint linear operator on~$V$, using~\eq{4.3}
we find that
$$
(a^\7 C a) f_S = \thalf H_{\comm CS} \w f_S.
$$

   Now from~\eq{5.2} we have
$$
2 \dm(X) f_S = H_{A_X} \w f_S + H_{\comm{C_X}S} \w f_S
 + \Tr_\C \lbrack SA_X\rbrack\, f_S - H_{SA_XS} \w f_S,
$$
so we arrive at
$$
\dm(X) f_S = \thalf(a^\7 A_X a^\7 + 2 a^\7 C_X a - a A_X a) f_S.
\eqno (6.3)
$$
Moreover, since $B(v) = a_J^\7(v) + a_J(v)$, it is readily
checked from the CAR that
$$
\thalf \comm{a^\7 A_X a^\7 + 2 a^\7 C_X a - a A_X a}{B(v)}
 = B(A_Xv + C_Xv) = B(Xv),
$$
so \eq{5.3} shows that \eq{6.3} holds with $f_S$ replaced by
$B(v)f_S$. We get, finally, for the current associated to~$X$:
$$
dG(X) = -\tihalf (a^\7 A_X a^\7 + 2 a^\7 C_X a - a A_X a),
\eqno (6.4)
$$
as an unbounded operator on the domain spanned by all $f_S$ and
$B(v)f_S$.

\subsection VI.2. Factorization of the scattering matrix

We interpret orthogonal transformations on~$V$ as classical
scattering transformations $S_{\rm cl}$. By ``classical'' we
mean the operator living in the one-particle space, to
distinguish it from the quantum scattering operator in Fock
space. But for a phase factor, $\mu(S_{\rm cl})$ is precisely
the $S$-matrix for a fermion system.

   We seek to factorize the operator $\mu(g)$ in a convenient
manner. Let us {\it define}, for $g \in SO'_*(V)$, the operators
$S_1$, $S_2$, $S_3$:
$$
S_1(g) = \exp(\thalf a^\7 T_g a^\7),  \ \quad
S_2(g) = \wick:\exp(a^\7(p_g^{-t} - I)a):\,,  \ \quad
S_3(g) = \exp(\thalf a \Hat T_g a).
$$
We compute the effect of these operators on Gaussians in a few
steps.

\proclaim Lemma 1.
$S_1$ is a bounded operator on~$\F(V)$, with
$S_1 f_R = f_{T_g + R}$ for any $R \in \Sk(V)$.

\Proof:
If $\Psi \in V^{\w k}$ for any finite~$k$, from~\eq{6.2}
we see that $(a^\7 T_g a^\7)^m \Psi = H_{T_g}^{\w m} \w \Psi$;
moreover, we have the norm estimate
$\norm{H_{T_g}^{\w m} \w \Psi} \leq \norm{H_{T_g}}^m\,\norm{\Psi}
 = \norm{T_g}_{\rm HS}^m \norm{\Psi}$, with
$\norm{T_g}_{\rm HS}$ denoting the Hilbert--Schmidt norm
of~$T_g$. Hence the series $\exp(\thalf a^\7 T_g a^\7)
 := \sum_{m=0}^\infty (2^m m!)^{-1} (a^\7 T_g a^\7)^m$ converges
in the bounded operator norm on~$\F_0(V)$, with the estimate
$\norm{\exp(\thalf a^\7 T_g a^\7)}
 \leq \exp(\thalf \norm{T_g}_{\rm HS})$.

   It is now immediate that $S_1\Psi = f_{T_g} \w \Psi$ for any
$\Psi \in \F(V)$. In particular, for $\Psi = f_R$, we have
$S_1 f_R = f_{T_g} \w f_R = f_{T_g + R}$.   \qed

\proclaim Lemma 2.
If $R \in \Sk(V)$ and $(I - R\Hat T_g)$ is invertible, then
$$
S_3 f_R = \det^{1/2}(I - R\Hat T_g) f_{R(I-\Hat T_gR)^{-1}}.
$$

\Proof:
This is straightforward, from
$\<f_S, S_3 f_R> = \<S_3^\7 f_S, f_R> = \<S_1(g^{-1})f_S, f_R>$.
\qed

\proclaim Lemma 3.
If $R \in \Sk(V)$, then $S_2 f_R = f_{p_g^{-t} R p_g^{-1}}$.

\Proof:
Firstly, if~$A$ is any bounded linear operator on~$(V,d,J)$,
then $ARA^t \in \Sk(V)$ and
$$
f_{ARA^t}
 = \sum_{K\rm\,finite} \Pf(R_K)\, Ae_{k_1}\w\cdots\w Ae_{k_{2m}};
  \qquad  K = \{k_1,\dots,k_{2m}\}.
\eqno (6.5)
$$
This is obtained from~\eq{3.3} on noting that
$H_{ARA^t} = \sum_{i,j} \<e_i,Re_j> Ae_i \w Ae_j$, which in turn
follows from the definition~\eq{3.1}.

   Secondly, we must check that, if $K_m = \{1,\dots,2m\}$, then
$$
\wick:\exp(a^\7Ca):\, \eps_{K_m}
 = (I+C)e_1 \w\cdots\w (I+C) e_{2m}.
\eqno (6.6)
$$
For then, if $K = \{k_1,\dots,k_{2m}\}$, we have
$\wick:\exp(a^\7Ca):\, \eps_K
 = (I+C)e_{k_1} \w\cdots\w (I+C) e_{k_{2m}}$ by a change of
orthonormal basis. Taking $C = p_g^{-t} - I$, \eq{6.5} gives the
result.

   To verify \eq{6.6}, note that the left hand side is a finite
series, since the terms $a(e_{l_j})$ with $l_j > 2m$ give no
contribution. Thus
$$
\eqalignno{
&\wick:\exp(a^\7Ca):\, \eps_{K_m}  \cr
&= \sum_{n=0}^{2m} {1\over n!}
 \sum_{\scriptstyle k_1\dots k_n \atop\scriptstyle l_1\dots l_n}
  a^\7(e_{k_1}) \dots a^\7(e_{k_n})
  \prod_{j=1}^n \<e_{k_j}, Ce_{l_j}>
  a(e_{l_n}) \dots a(e_{l_1}) \eps_{K_m}  \cr
&= \sum_{L\subset K_m} {1\over \abs L!} \eta_L \,
  a^\7(Ce_{l_1}) \dots a^\7(Ce_{l_n}) \eps_{K_m\setminus L}
 = \sum_{L\subset K_m} f_1 \w\cdots\w f_{2m},
\cr}
$$
where $\eta_L$ is the sign of the permutation
$K_m\mapsto (L,K_m\setminus L)$ and $f_j = Ce_j$ if~$j \in L$,
$f_j = e_j$ otherwise. But the latter sum is just an expansion
of the right hand side of~\eq{6.6}.   \qed

\smallskip

   It is worth noting that since
$$
f_{p_g^{-t} R p_g^{-1}}
 = \sum_K \Pf((p_g^{-t} R p_g^{-1})_K) \eps_K
 = \sum_K \detc((p_g^{-1})_K) \Pf(R_K) \eps_K
$$
and $\abs{\detc((p_g^{-1})_K)}
 = \det^{1/2}((p_g^{-t})_K (p_g^{-1})_K)
 \leq \det^{1/2}(p_g^{-t} p_g^{-1}) = \det^{-1/2}(I - T_g^2)$,
then $S_2$ extends to~$\F_0(V)$ as a bounded operator with norm
at most $c_g^2$.

\smallskip

   Now we see that, on applying $S_1$, $S_2$, $S_3$ in turn
to~$f_R$, the index of the Gaussian transforms as
$R \mapsto T_g + p_g^{-t} R(I - \Hat T_gR)^{-1} p_g^{-1} = g\.R$
by~\eq{4.4}, and so by~\eq{4.5}:
$$
c_g S_1(g) S_2(g) S_3(g)\, f_R
 = c_g \det^{1/2}(I - R\Hat T_g) f_{g\.R} = \mu(g) f_R,
\eqno (6.7)
$$
whenever $p_g^{-1}$ and~$g\.R$ exist. Thus
$\mu(g) = c_g S_1 S_2 S_3$ holds on~$\F_0(V)$ whenever
$g \in SO'_*(V)$.

  Finally, if $p_g$ is not invertible, the general factorization
is obtained from
$$
\mu(g) = c_{rg}\, B(e_1) \dots B(e_n)\, S_1(rg) S_2(rg) S_3(rg)
\eqno (6.8)
$$
where $n = \dim(\ker p_g)$.

\section VII. Quantization of the Dirac equation

\subsection VII.1. The choice of complex structures

We examine only the case of a charged field. Thus we think
of~$V$ as the space of complex solutions of the free Dirac
equation:
$$
i{\partial \over \partial t} \psi
 = (\alb\.\pb + \beta m) \psi =: H\psi,
$$
where $\pb = -i\,\partial/\partial\xb$, regarded as a
{\it real\/} vector space with the symmetric form:
$$
d(\psi_1,\psi_2) = \half\biggl( \int \psi_1^*\psi_2 \,d^3x
   + \int \psi_2^*\psi_1 \,d^3x\biggr).
$$

   For definiteness, we shall adopt the Dirac representation of
the $\alb$ and~$\beta$ matrices:
$$
\alb = \pmatrix{0 & \sgb \cr \sgb & 0 \cr};  \qquad
\beta = \pmatrix{I & 0 \cr 0 & -I \cr}.
$$
The operator $H$ is selfadjoint with domain
$\Dom(H) \subset \H := \C^4 \otimes L^2(\R^3)$. Define the
two-spinor functions:
$$
u_s(\kb) := \pmatrix{\sqrt{\omega(\kb) + m}\, \chi_s \Strut\cr
  {\displaystyle{\sgb\.\kb \over \sqrt{\omega(\kb) + m}}}\,
  \chi_s \Strut\cr};
\qquad
v_s(\kb) := \pmatrix{{\displaystyle{\sgb\.\kb
               \over \sqrt{\omega(\kb) + m}}}\, \chi_s \Strut\cr
             \sqrt{\omega(\kb) + m}\, \chi_s \Strut\cr};
$$
where $s = \sube$ or~$\baja$, as the case may be,
$\chi_\sube = {1\choose0}$ and $\chi_\baja = {0\choose1}$.
Denote by $(\.,\.)$ the ordinary hermitian product on~$\C^2$.
Then one checks that:
$$
\displaylines{
\bigl( u_s(\kb), u_{s'}(\kb) \bigr)
 = \bigl( v_s(\kb), v_{s'}(\kb) \bigr)
 = 2 \omega(\kb) \delta_{ss'},  \cr
\bigl( u_s(\kb), v_{s'}(-\kb) \bigr) = 0.
\cr}
$$

   We consider also the projectors:
$$
P_\pm = \thalf (1 \pm (\alb\.\pb + \beta m) \omega^{-1} ),
$$
corresponding respectively to the positive and negative parts of
the spectrum of~$H$. Then $P_+ + P_- = I$ and
$H P_\pm = \pm \omega P_\pm$; we may write $V_\pm = P_\pm V$. We
note also the relations:
$$
\eqalign{P_+(\kb) u_s(\kb) &= u_s(\kb), \cr
         P_-(\kb) u_s(\kb) &= 0, \cr}  \qquad
\eqalign{P_+(\kb) v_s(-\kb) &= 0, \cr
         P_-(\kb) v_s(-\kb) &= v_s(-\kb), \cr}
\eqno (7.1)
$$
where we have denoted by $P_\pm(\kb)$ the projectors on the
Fourier transformed space of~$\H$, which are multiplication
operators. Moreover,
$$
\eqalign{
u_\sube(\kb) u_\sube^\7(\kb) + u_\baja(\kb) u_\baja^\7(\kb)
&= 2 \omega(\kb) P_+(\kb), \cr
v_\sube(-\kb) v_\sube^\7(-\kb) + v_\baja(-\kb) v_\baja^\7(-\kb)
&= 2 \omega(\kb) P_-(\kb). \cr}
\eqno (7.2)
$$

   Besides $\H$, we shall consider the Hilbert space $\Onda\H$,
which is~$V$ endowed with the new complex Hilbert space
structure given by $d(\.,\.) + id(J\.,\.)$, with
$J := i(P_+ - P_-)$. In other words, we preserve the real
part~$d$ of the inner product and we introduce a nonlocal
imaginary part through the nonlocal complex structure. In this
way, complex multiplication is dynamically built into the space
of solutions, in such a manner to make possible a direct
interpretation of ``negative energy'' solutions as antiparticles.

   Now define the 2-spinor functions:
$$
b(\kb)
 := \pmatrix{\bigl( u_\sube(\kb), \F\psi(\kb) \bigr) \Strut\cr
      \bigl(u_\baja(\kb), \F\psi(\kb) \bigr) \Strut\cr},  \qquad
d(\kb)
 := \pmatrix{\bigl(\F^{-1}\psi(\kb), v_\sube(\kb)\bigr) \Strut\cr
      \bigl(\F^{-1}\psi(\kb), v_\baja(\kb)\bigr) \Strut\cr}.
$$
In view of~\eq{7.1} and~\eq{7.2}, this transformation is inverted
by:
$$
\psi(\xb) = (2\pi)^{-3/2} \sum_{s=\sube,\baja} \int
 (b_s(\kb) u_s(\kb) e^{i\kb\xb}
  + d_s^\7(\kb) v_s(\kb) e^{-i\kb\xb}) \,d\mu(k),
$$
where $d\mu(k) := d^3k/\omega(\kb)$. It is seen now that
the map to momentum space
$\Onda\H \to \H_m^{\shalf,+} \oplus \H_m^{\shalf,+}$, where
$\H_m^{\shalf,+} = \C^2 \otimes L^2(H_m^+,d\mu)$, given by
$\psi \mapsto {b \choose d}$, is an
isometry such that $J\psi \mapsto i{b \choose d}$. Here we make
contact with Weinberg's quantization method,\cite{\Weinberg}
based on Wigner's theory of the unitary irreducible
representations of the Poincar\'e group. It will follow from our
treatment in the next subsection that the fermion field has the
gauge transformation properties required in the Weinberg
construction.

\subsection VII.2. Quantization and the charge operator

We now prove that the {\it standard\/} construction of fermion
Fock space of Section~III, applied to~$\Onda\H$, or
$\H_m^{\shalf,+} \oplus \H_m^{\shalf,+}$, gives the charged
fermion field. In order to fully grasp what is involved here,
we need some further reflection on the relation between the two
Hilbert space structures for the space of solutions of the Dirac
equation. We can define the charge operator on the one-particle
space for the Dirac equation as the generator of gauge
transformations: $Q\psi := i\psi$. It is an infinitesimally
orthogonal operator:
$$
d(\psi_1, Q\psi_2) + d(Q\psi_1, \psi_2) = 0.
$$
By definition we have $(V,d,Q) \equiv \H$ and
$(V,d,J) \equiv \Onda\H$. One can pass from the ``natural''
complex structure $Q$ to $J$ by means of an orthogonal
transformation of~$V$:
$$
g_0 Q g_0 = J,  \sepword{with}
  g_0 := \pmatrix{I & 0 \cr 0 & C \cr} = g_0^{-1},
\eqno (7.3)
$$
in the $V_+ \oplus V_-$ splitting. Thus the Dirac equation
on~$\H$:
$$
i{\partial\over\partial t}\psi = H\psi
$$
becomes:
$$
J{\partial\over\partial t}\psi = -iJH\psi,
$$
the transformed operator $\Onda H := -iJH$ being selfadjoint on
$\Onda\H$. Obviously $-iH$ and $-J\Onda H$ are the same element
of the orthogonal Lie algebra (in the generalized sense, since
these are unbounded operators). The crucial difference is that
$\Onda H$ is bounded below, in fact positive:
$$
\Onda HP_\pm = \omega P_\pm.
$$
The orthogonal transformation
$g_0$ has changed the spectrum! The idea is apparently due to
Bongaarts, although its implementation in~Ref.~\Bongaarts\ is
rather murky; one should look also at~Refs.~\Saunders\ and
\Thaller. As $g_0$ does not fulfil the Shale--Stinespring
criterion, $\F(\H)$ and $\F(\Onda\H)$ are not equivalent upon
quantization. We must choose $\Onda\H$ for the {\it standard\/}
construction of fermion Fock space ---which allows the
straightforward particle interpretation--- to yield the charged
fermion field. This is a crucial point in our argument, because
otherwise we would not be able to use the spin representation in
QED.

   In view of~\eq{7.3}, we can apply the Bogoliubov
transformation philosophy to relate creation and annihilation
operators defined with respect to each complex structure. Let us
abbreviate $\psi_+ := P_+\psi$, $\psi_- := P_-\psi$. Then
$p_{g_0}\psi = \psi_+$, $q_{g_0}\psi = \psi_-^*$. We get on
$\F(\H)$, from~\eq{4.2}:
$$
a_Q(g_0\psi)    = a_J(\psi_+) + a_J^\7(\psi_-^*); \qquad
a_Q^\7(g_0\psi) = a_J^\7(\psi_+) + a_J(\psi_-^*),
$$
so that
$$
a_Q(\psi)    = a_J(\psi_+) + a_J^\7(\psi_-); \qquad
a_Q^\7(\psi) = a_J^\7(\psi_+) + a_J(\psi_-)
$$
and similar converse equations in $\F(\Onda \H)$.

Note also the simple formula:
$$
\<\psi,\phi>_{\scriptscriptstyle J}
 = \bbraket{\psi_+}{\phi_+} + \bbraket{\phi_-}{\psi_-},
\eqno (7.4)
$$
where $\bbraket\.\.$ will now denote the ``natural'' inner
product $\<\.,\.>_{\scriptscriptstyle Q}$ of~$\H$.

   Now we prove ---for operators on $V$ preserving both
structures--- that our quantization method gives the same result
as the usual procedure of first performing the Fock quantization
\wrt~the ``natural'' complex structure and then amending the
result with a ``normal ordering'' recipe; the tricky subtraction
of infinities is avoided.

\proclaim Proposition.
If $X$ commutes with both~$J$ and~$Q$, then
$dG(X) = \wick:d\Gamma_Q(-iX):\,$.

\Proof:
 From~\eq{6.4} we have
$dG(X) = d\Ga_J(-JX) = -i\, a_J^\7 X a_J^{\vphantom{\7}}$ since
$XJ = JX$. Let us choose orthonormal bases $\{\varphi_k\}$ and
$\{\psi_k\}$ for~$V_+$ and~$V_-$. We write $b^{(\7)}$ rather
than $a^{(\7)}$ on~$V_+$, $d^{(\7)}$ rather than $a^{(\7)}$
on~$V_-$, as is customary. If $Y = -JX$, then $Y$ is selfadjoint
on~$\Onda\H$, and so
$$
\eqalignno{
&-i\, a_J^\7 X a_J^{\vphantom{\7}}
 =\sum_{j,k} b_J^\7(\varphi_k)
  \<\varphi_k, Y\varphi_j>_{\scriptscriptstyle J} b_J(\varphi_j)
  + d_J^\7(\psi_k)
    \<\psi_k, Y\psi_j>_{\scriptscriptstyle J} d_J(\psi_j) \cr
&=\sum_{j,k} b_J^\7(\varphi_k) \bbraket{\varphi_k}{-iX\varphi_j}
    b_J(\varphi_j) - d_J^\7(\psi_k) \bbraket{\psi_j}{-iX\psi_k}
    d_J(\psi_j)
 = \wick:d\Gamma_Q(-iX):\,.
& \qed   \cr}
$$

\smallskip

   In particular, we have:
$$
\Q := dG(Q) = b^\7b - d^\7d
$$
in units of electronic charge. This current has integer
eigenvalues and we call charge sectors the eigenspaces
$\F_k(V)$, for $k\in\Z$. We remark that the 1-particle charge
conjugation operator, which is antilinear in $\H$, is linear in
$\Onda\H$.\cite{\Saunders,\Thaller}

   The particle interpretation of the quantum field and the
possibility of a direct {\it translation into physics of our
results on the spin representation\/} hinge on our choice of $J$
as complex structure, one that allows us to dry up the Dirac sea
---which is what the physical vacuum looks like using the
``natural'' (and wrong) complex structure. However, as local
conservation of the charge in interactions is a basic physical
principle, we cannot dispense entirely with $Q$ in the
quantization process; the interplay of both complex structures
is characteristic of the theory of charged fields. This is
reflected in the fact that the invariance group of the theory is
not $O'(V)$, but its subgroup $U'_Q(V)$ of (restricted) unitary
operators on $H$, which has a very different topological
structure: whereas $O'(V)$ has two connected pieces, we will soon
see that the group $U'_Q(V)$ has an infinite number of connected
pieces, naturally indexed by $\Z$.

\subsection VII.3. Charge sectors

Now we are prepared to translate the group-representation
machinery into the usual language for QED. We relabel the vacuum
vector $\Omega$ as $\vac$, and write $\vacout := \mu(g)\vac$ for
the out vacuum. From~\eq{3.2} and~\eq{6.2}, it follows at once
that
$$
\exp(\thalf a^\7 T a^\7) \vac = f_T
$$
for $T \in \Sk(V)$. Thus $\vacout$ is proportional to
$\exp(\thalf a^\7 T_g a^\7) \vac$ whenever $g \in SO'_*(V)$.

   If the classical scattering operator $g = S_{\rm cl}$ lies
in~$O'(V)$ but with $\dim(\ker p_g) = n > 0$, we again write
$g = r_{e_1}\dots r_{e_n}h$, where $\{e_1,\dots,e_n\}$ is an
orthonormal basis of $\ker p_g^t$; the out-vacuum can thus be
written as
$$
\vacout \propto e_1 \w\cdots\w e_n \w f_{T_{rg}}
 = a^\7(e_1) \dots a^\7(e_n) \exp(\thalf a^\7 T_{rg} a^\7) \vac.
\eqno (7.5)
$$

   In this subsection and most of what follows, we shall always
assume that a particle-antiparticle fermion field can be built
over the one-particle space $V$, and that we always have
$J = i(P_+ - P_-)$, where $P_\pm = I-P_\mp$ denote orthogonal
projectors on infinite dimensional subspaces, the outstanding
example being the space of solutions of a Dirac equation. This
is to say, we propose to deal with charged fermion fields; even
so, all we have to say in the next subsection is also valid for
neutral fields.

   We write then $g$, $p_g$, $q_g$, $T_g$, $\Hat T_g$ in
matricial form, with respect to the decomposition
$V = V_+ \oplus V_-$, with the proviso that $T_g$ exists iff
$p_g$ is invertible:
$$
g = \pmatrix{S_{++} & S_{+-} \cr S_{-+} & S_{--} \cr},
    \sepword{thus}
p_g = \pmatrix{S_{++} & 0 \cr  0 & S_{--}}, \quad
q_g = \pmatrix{0 & S_{+-} \cr S_{-+} & 0}.
$$
The fact that $g \in U'_Q(V)$ means precisely (as remarked at
the end of the previous Section) that the $S_{\pm\pm}$ are
{\it complex-linear\/} operators acting between the complex
spaces $V_+$, $V_-$. We leave to the care of the reader to
rewrite \eq{2.1} in terms of the $S$'s. It is immediate that
$$
T_g
 = \pmatrix{0 & S_{+-}S_{--}^{-1}\cr S_{-+}S_{++}^{-1} & 0\cr},
\quad
\Hat T_g
 = \pmatrix{0 & -S_{++}^{-1}S_{+-}\cr -S_{--}^{-1}S_{-+} & 0\cr}.
$$
We see that $p_g^{-1}$ exists if and only if $S_{++}$ and
$S_{--}$ are invertible, and that $g \in O'(V)$ iff $S_{+-}$
and~$S_{-+}$ are Hilbert--Schmidt (actually, since
$S_{+-}(g^{-1}) = (S_{-+}(g))^\7$, it suffices that $S_{+-}$ be
Hilbert--Schmidt).

   From the fact, remarked in Section~II, that $\ind p_g = 0$,
we get $\ind S_{++} = -\ind S_{--}$. One checks that
$g(\ker S_{\pm\pm}) = \ker S_{\mp\mp}^\7$, and so
$\ind S_{\pm\pm}
 = \dim\ker S_{\mp\mp}^\7 - \dim\ker~S_{\pm\pm}^\7$.

\smallskip

   Now, the $f_{T_g}$ are all charge zero states, as
$\gamma(g,Q) = 0$ for all $g \in SO'_*(V) \cap U'_Q(V)$.

\smallskip

   For out-vacua of the form~\eq{7.5}, we can always choose
orthonormal basis $\{\varphi_1,\dots,\varphi_l\}$ and
$\{\psi_1,\dots,\psi_m\}$ for~$\ker S_{++}^\7$ and
$\ker S_{--}^\7$ respectively, so that $l + m = n$. We have
clearly, for the expectation value of the charge in the out
vacuum $\vacout = \mu(g)\vac$:
$$
\braCket{0_{\rm out}}{\Q}{0_{\rm out}} = l - m
 = \dim\ker S_{++}^\7 - \dim\ker S_{--}^\7
 = \ind S_{--} = -\ind S_{++},
$$
and then:
$$
\mu(g)\F_k(V) = \F_{k+\ind S_{--}}(V),
$$
which can be rewritten as
$$
\mu(g)\Q\mu(g)^{-1} = \Q + \ind S_{++}.
\eqno (7.6)
$$
Thus the group $U'_Q(V)$ has infinitely many connected
components, indexed by the Fredholm index of $S_{++}$ or
$S_{--}$; which components interchange the charged vacua. Note
that~\eq{7.6} is an {\it anomalous\/} identity, as $Q$ commutes
both with $g$ and~$J$. However, it has been shown by Carey and
O'Brien\cite{\Careyob} in QED and then for quite general gauge
fields by Matsui\cite{\Matsui} that under reasonable
circumstances the scattering matrix belongs to the component of
the identity $U'_{Q,0}(V)$ of the group; thus vacuum
polarization in this sense does not occur in the external field
problem ---where consequently only pair creation arises.

   A similar treatment is possible for the chiral charge
anomaly. Let us consider, for simplicity, a theory of massless
fermions in $1+1$ spacetime dimensions. Then again
$Q_5 := i\gamma_5$ is an infinitesimally orthogonal operator,
commuting with $J$, which takes in momentum space the following
form, with respect to the $V = V_+ \oplus V_-$ splitting:
$$
Q_5 = \pmatrix{i\epsilon(k) & 0 \cr 0 & i\epsilon(-k) \cr}.
$$
The support in momentum space of the elements of an orthonormal
basis for $p_g^t$ must now lie either in the right or the left
half axis. With an obvious notation, we have for the chiral
current:
$$
\Q_5 := dG(Q_5) = b_R^\7b_R - b_L^\7b_L + d_L^\7d_L - d_R^\7d_R.
$$
Again $\Q_5$ has integer eigenvalues and (for scattering
operators $g$ such that $\comm{g}{Q_5} = 0$) a formula of the
type~\eq{7.6} intervenes; only now the index of the scattering
operator is directly related to the Chern number of the gauge
field\cite{\Matsui} and it is generally nonzero for elements of
$U'_{Q,0}(V)$. This is why the chiral charge anomaly is local in
the usual parlance.\cite{\AGaume} Index formulae for currents
associated with special gauge transformations are given
in~Ref.~\Ruijsenaarsbis. Further consideration of these
questions would take us too far afield.

\subsection VII.4. The scattering matrix
                   for a charged fermion field

For simplicity, we will assume for the rest of this subsection
that $p_g$ is invertible. We need the formulae:
$$
\Tr_\C \pmatrix{A_{++} & 0 \cr 0 & A_{--} \cr}
 = \Tr(A_{++}) + \Tr(A_{--}^\7)
\eqno (7.7)
$$
and by exponentiating:
$$
\detc \pmatrix{A_{++} & 0 \cr 0 & A_{--} \cr}
 = \det(A_{++})\, \det(A_{--}^\7).
\eqno (7.8),
$$
which come from~\eq{7.4}. For instance, using~\eq{7.7} and since
$X_{+-}^\7 = - X_{-+}$, the Schwinger terms~\eq{5.7}
reduce to:
$$
\eqalignno{
\comm{dG(X)}{dG(Y)} &= -\a(X,Y)
 = \thalf \Tr(X_{+-}Y_{-+} - Y_{+-}X_{-+})
 + \thalf \Tr(Y_{-+}X_{+-} - X_{-+}Y_{+-})  \cr
&= \Tr(X_{+-}Y_{-+} - Y_{+-}X_{-+}) = 2i\Im\Tr(X_{+-}Y_{-+}).
\cr}
$$
The fermionic anomaly $\ga(g,X)$ can also be recomputed in the
QED language. From~\eq{5.10} and~\eq{7.7} one gets:
$$
\eqalignno{
\ga(g,X)
&= \Tr\bigl( (I - S_{++}^{-1}S_{+-}S_{--}^{-1}S_{-+})^{-1}
 (X_{+-}S_{--}^{-1}S_{-+} - S_{++}^{-1}S_{+-}X_{-+}  \cr
&\hskip 6em - S_{++}^{-1}S_{+-}S_{--}^{-1}S_{-+}X_{++}
        + S_{++}^{-1}S_{+-}X_{--}S_{--}^{-1}S_{-+}) \bigr). \cr}
$$

\smallskip

   For the absolute value of the vacuum persistence amplitude we
obtain, since $T_g$ is skewsymmetric:
$$
\eqalignno{
\abs{\vacpersamp} &= \det^{-1/4}(I - T_g^2)
 = \det^{-1/2}(I + (S_{+-} S_{--}^{-1})^\7 S_{+-} S_{--}^{-1})\cr
&= \det^{-1/2}((S_{--}^\7)^{-1} (S_{--}^\7 S_{--}
                + S_{+-}^\7 S_{+-}) S_{--}^{-1})  \cr
&= \det^{-1/2}((S_{--}^\7)^{-1} S_{--}^{-1})
 = \det^{1/2}(S_{--}S_{--}^\7), \cr}
$$
using $S_{--}^\7 S_{--} +  S_{+-}^\7 S_{+-} = I$ (on $V_-$). On
the other hand,
$$
\abs{\vacpersamp} = \det^{1/4}(p_g p_g^t)
 = \det^{1/4}(S_{++} S_{++}^\7) \,\det^{1/4}(S_{--} S_{--}^\7),
$$
and hence both factors on the right hand side are equal. We thus
arrive at
$$
\eqalignno{
\abs{\vacpersamp} &= \det^{1/2}(S_{--} S_{--}^\7)
 = \det^{1/2}(S_{++} S_{++}^\7)  \cr
&= \det^{1/2}(I - S_{+-} S_{+-}^\7)
 = \det^{1/2}(I - S_{-+} S_{-+}^\7). \cr}
$$

\smallskip

   We are finally ready for the computation of the full
$S$-matrix. We start from the factorization~\eq{6.7}. We choose
orthonormal bases $\{\varphi_k\}$ and $\{\psi_k\}$ for~$V_+$
and~$V_-$ and regard their union as an orthonormal basis for
$V$. We distinguish the particle and antiparticle sectors by
setting $b^\7(\varphi_k) := a_J^\7(\varphi_k)$,
$d^\7(\psi_k) := a_J^\7(\psi_k)$, \dots\  Then we find:
$$
\eqalignno{
\thalf a^\7 T_g a^\7
&= \half\sum_{j,k} a_J^\7(\varphi_k)
     \<\varphi_k, T_g \psi_j> a_J^\7(\psi_j) + a_J^\7(\psi_j)
     \<\psi_j, T_g \varphi_k> a_J^\7(\varphi_k) \cr
&= \half\sum_{j,k} b^\7(\varphi_k) \bbraket{\varphi_k}{T_g\psi_j}
     d^\7(\psi_j) + d^\7(\psi_k) \bbraket{T_g\varphi_j}{\psi_k}
     b^\7(\varphi_j)  \cr
&= \half \sum_{j,k} b^\7(\varphi_k)
     \bbraket{\varphi_k}{S_{+-}S_{--}^{-1}\psi_j} d^\7(\psi_j)
   + d^\7(\psi_j) \bbraket{S_{-+}S_{++}^{-1}\varphi_k}{\psi_j}
     b^\7(\varphi_k) \cr
&= \sum_{j,k} b^\7(\varphi_k)
    \bbraket{\varphi_k}{S_{+-}S_{--}^{-1}\psi_j} d^\7(\psi_j)
 =: b^\7 S_{+-}S_{--}^{-1} d^\7,
& (7.9)  \cr}
$$
using the CAR $\{b^\7(\varphi_j),d^\7(\psi_k)\} = 0$, the
relation $(S_{-+}S_{++}^{-1})^\7 = -(S_{+-}S_{--}^{-1})$,
and~\eq{7.4}. In like manner, we obtain
$$
\eqalignno{
\thalf a \Hat T_g a
&= \half \sum_{j,k} a_J(\varphi_k)
     \<\Hat T_g \varphi_k, \psi_j> a_J(\psi_j) + a_J(\psi_j)
     \<\Hat T_g \psi_j, \varphi_k> a_J(\varphi_k) \cr
&= \half\sum_{j,k} b(\varphi_k)
     \bbraket{\psi_j}{\Hat T_g \varphi_k} d(\psi_j) + d(\psi_j)
     \bbraket{\Hat T_g\psi_j}{\varphi_k} b(\varphi_k)  \cr
&= - \half \sum_{j,k} b(\varphi_k)
     \bbraket{\psi_j}{S_{--}^{-1}S_{-+}\varphi_k} d(\psi_j)
   + d(\psi_j) \bbraket{S_{++}^{-1}S_{+-}\psi_j}{\varphi_k}
     b(\varphi_k) \cr
&= \sum_{j,k} d(\psi_j)
     \bbraket{\psi_j}{S_{--}^{-1}S_{-+}\varphi_k}
     b(\varphi_k) =: d S_{--}^{-1}S_{-+} b.
& (7.10)  \cr}
$$

   The Wick-ordered product $\wick:\exp( a^\7(p_g^{-t} - I)a ):$
can be written as $S_{2b} S_{2d}$ by separating the $b^{(\7)}$
and $d^{(\7)}$ terms. Since
$\bbraket{\varphi_k}{(p_g^{-t} - I)\varphi_l}
 = \bbraket{\varphi_k}{((S_{++}^\7)^{-1} - I)\varphi_l}$, we have
$S_{2b} = \wick:\exp( b^\7((S_{++}^\7)^{-1} - I)b ):\,$, and for
$S_{2d}$ we get
$$
\eqalignno{
S_{2d}
&= \sum_{n=0}^\infty {1\over n!}
 \sum_{\scriptstyle k_1\dots k_n \atop\scriptstyle l_1\dots l_n}
 d^\7(\psi_{k_1}) \dots d^\7(\psi_{k_n})  \prod_{j=1}^n
 \bbraket{\psi_{l_j}}{(I - S_{--}^{-1})\psi_{k_j}}
 d(\psi_{l_n}) \dots d(\psi_{l_1})    \cr
&= \wick:\exp( d(I - S_{--}^{-1})d^\7 ):\,.
& (7.11)  \cr}
$$

    Putting the equations \eq{7.9--11} together, we arrive at
the exact $S$-matrix for the charged fermion field:
$$
\eqalignno{
{\bm S} = e^{i\theta} \mu(g)
&= \vacpersamp\, \exp(b^\7 S_{+-}S_{--}^{-1} d^\7)
& (7.12)  \cr
&\quad\x \wick:\exp(b^\7((S_{++}^\7)^{-1} - I)b
  - d(S_{--}^{-1} - I)d^\7):\, \exp(d S_{--}^{-1}S_{-+} b).
\cr}
$$
Whenever $S_{++}$ and $S_{--}$ are not invertible, this formula
must be modified in accordance with~\eq{6.8}.

   It is instructive to compare the form~\eq{7.12} of the
fermionic $S$-matrix with the bosonic $S$-matrix expression
obtained in a parallel way from the metaplectic
representation.\cite{\Aldebaran} We remark that the exact
quantum $S$-matrix for the external field problem has direct
application in laser physics and heavy ion collisions.

   Let $\varphi\in V_+$ and $\psi\in V_-$. The following
commutation rules (and their adjoints) are very useful for
calculations involving the $S$-matrix:
$$
\eqalignno{
\comm{e^{b^\7Ad^\7}}{b(\varphi)}
 = -d^\7(A^\7\varphi) e^{b^\7Ad^\7}, \quad
&\quad
\comm{e^{b^\7Ad^\7}}{d(\psi)} = b^\7(A\psi) e^{b^\7Ad^\7},
& (7.13)  \cr
\wick:e^{b^\7(A-I)b}:\, b^\7(\varphi)
 =  b^\7(A\varphi)\, \wick:e^{b^\7(A-I)b}:\,,  \quad
&\quad
\wick:e^{d(I-A)d^\7}:\, d^\7(\psi)
 = d^\7(A^\7\psi)\, \wick:e^{d(I-A)d^\7}:\,.
\cr}
$$

\section VIII. The Feynman rules for electrodynamics
               of external fields

We next derive the Feynman rules for quantum electrodynamics of
external fields from the exact $S$-matrix. The Dirac equation
in an external electromagnetic field is
$$
i {\partial \over \partial t} \psi = (H + V)\psi,
$$
where the external field $V$ is given by
$$
V\psi = e (-\alb\.{\bm A} + A_0) \psi = e \ga^0 \whack A \psi,
$$
with the usual notation $\whack A := \ga^\mu A_\mu$. The
classical scattering matrix corresponding to this problem is
given by a Dyson expansion:
$$
S_{\rm cl}
 = \sum_{n=0}^\infty (-i)^n \int_{-\infty}^\infty
    \int_{-\infty}^{t_1} \cdots \int_{-\infty}^{t_{n-1}}
     \Onda V(t_1) \dots \Onda V(t_n) \,dt_n \dots \,dt_2 \,dt_1
 =: \sum_{n=0}^\infty S_{:n}.
$$
Here $\Onda V(t) := e^{iHt} V(t) e^{-iHt}$. We write out $S_{:n}$
as an integral kernel in momentum space:
$$
\eqalignno{
S_{:n}(\kb,\kb')
&:= {(-i)^n\over (2\pi)^{3n/2}} \int_{-\infty}^\infty \cdots
    \int_{-\infty}^\infty \int \cdots \int
    e^{iH(\kb)t_1} V(t_1, \kb-\kb_1) \cr
&\quad\x\Strut \theta(t_1 - t_2) e^{-iH(\kb_1)(t_1-t_2)}
    V(t_2, \kb_1-\kb_2) \dots \theta(t_1 - t_n)
    e^{-iH(\kb_{n-1})(t_{n-1}-t_n)} \cr
&\quad\x\Strut V(t_n, \kb_{n-1} - \kb') e^{-iH(\kb')t_n}
  \,d^3k_1 \dots \,d^3k_{n-1} \,dt_n \dots \,dt_1,  \cr}
$$
where $\theta$ denotes the Heaviside function.

   We shall not dwell on the question of the conditions on $V$
such that $S_{\rm cl}$ is implementable. An apparently more
ambitious endeavour would be to try to implement the interacting
time evolution operator $U(t,t')$ with
$U(\infty,-\infty) = S_{\rm cl}$. It can be shown, however, that
$U(t,t')$ can be implemented only for electric fields; hence the
implementability of time evolution has no covariant meaning and
the particle interpretation has meaning only in the realm of
scattering theory.

   To derive the Feynman rules, the first step is to rewrite
$S_{:n}$ in covariant form. We follow the treatment
of~Ref.~\Scharf. From the formula:
$$
\theta(t)  e^{-iH(\kb)t}
 = {i \over 2\pi} \int dk^0 S_\ret(k) \ga^0  e^{-ik^0t},
$$
where
$$
S_\ret(k) := {\whack{\,k} + m \over k^2 - m^2 + ik^0 0},
$$
we obtain:
$$
\eqalignno{
(S_{\pm\pm})_{:n}(\kb,\kb')
&= -i(2\pi)^{-2n+1} e^n P_\pm(\kb) \ga^0
   \biggl( \int \cdots \int \whack A(k - k_1) S_\ret(k_1)  \cr
&\quad\x{} \whack A(k_1 - k_2) \dots S_\ret(k_{n-1})
   \whack A(k_{n-1} - k') \,d^4k_1 \dots \,d^4k_{n-1} \biggr)
   P_\pm(\kb')  \cr}
$$
(where $k^0 = \omega(\kb)$ and ${k'}^0 = \omega(\kb')$ are
understood).

\smallskip

   It is well known and physically obvious ---as beautifully
discussed in the classic paper Ref.~\Feynman--- that the
$n$-pair amplitudes are Slater determinants of the one-pair
amplitudes. It is enough thus to derive the one-pair amplitudes.
There are four of them, which are not altogether independent;
their expressions may be computed from~\eq{7.12} and the rules
\eq{7.13} of commutation of the creation and annihilation
operators with the quadratic exponentials.

\smallskip

\item{1.}
For electron scattering from initial state~$\varphi_i$ to final
state~$\varphi_f$:
$$
S_{fi}
:= \<b^\7(\varphi_f)0_{\rm in}, {\bm S}b^\7(\varphi_i)0_{\rm in}>
 = \vacpersamp \, \bbraket {S_{++}^{-1} \varphi_f} {\varphi_i};
$$

\item{2.}
For positron scattering from initial state~$\psi_i$ to final
state~$\psi_f$:
$$
S_{fi}
 := \<d^\7(\psi_f) 0_{\rm in}, {\bm S}d^\7(\psi_i) 0_{\rm in}>
 = \vacpersamp \, \bbraket {\psi_i} {S_{--}^{-1} \psi_f};
$$

\item{3.}
For creation of an electron-positron pair, in respective states
$\varphi$, $\psi$:
$$
S_{fi}
 := \<b^\7(\varphi) d^\7(\psi) 0_{\rm in}, {\bm S} 0_{\rm in}>
 = \vacpersamp \, \bbraket {\varphi} {S_{+-} S_{--}^{-1} \psi};
$$

\item{4.}
For annihilation of an electron-positron pair, in respective states
$\varphi$, $\psi$:
$$
S_{fi}
 := \<0_{\rm in}, {\bm S} b^\7(\varphi) d^\7(\psi) 0_{\rm in}>
 = \vacpersamp \, \bbraket {\psi} {S_{--}^{-1} S_{-+} \varphi}.
$$

\noindent
Note that if $V$ is time-independent, by a well-known result of
scattering theory, one has $\comm{S_{\rm cl}}{H} = 0$; thus
$\comm{S_{\rm cl}}{J} = 0$ and there cannot be creation or
annihilation of pairs. In such a context, the quantized and the
one-particle theory are essentially equivalent.

\smallskip

   We need to compute an expansion for $S_{\pm\pm}^{-1}$ from
the expansion of~$S_{\rm cl}$, in order to proceed. From the
identity $I + \sum_{n\geq1} (S^{-1})_{:n} = S^{-1}
 = (I + \sum_{n\geq1} S_{:n})^{-1}$, we have:
$$
(S^{-1})_{:n} = -\bigl( S_{:n} + S_{:n-1}(S^{-1})_{:1}
 + S_{:n-2}(S^{-1})_{:2} +\cdots+ S_{:1}(S^{-1})_{:n-1} \bigr),
\eqno (8.1)
$$
where juxtaposition means convolution of kernels:
$ST(\kb,\kb') := \int S(\kb,\kb_1) \,T(\kb_1,\kb') \,d^3k_1$. For
instance:
$$
\eqalignno{
&(S_{++}^{-1})_{:1} (\kb,\kb') = - (S_{++})_{:1} (\kb,\kb')
 = {i\over2\pi}e\, P_+(\kb) \ga^0 \whack A(k-k') P_+(\kb');  \cr
&(S_{++}^{-1})_{:2} (\kb,\kb')
 = - (S_{++})_{:2} (\kb,\kb') + \int (S_{++})_{:1} (\kb,\kb_1)
     (S_{++})_{:1} (\kb_1,\kb') \,d^3k_1 \cr
&\quad = i(2\pi)^{-3} e^2\, P_+(\kb) \ga^0
   \biggl( \int \whack A(k - k_1) S_\ret(k_1)
     \whack A(k_1 - k') \,d^4k_1 \biggr) P_+(\kb') \cr
&\qquad - (2\pi)^{-2} e^2\, P_+(\kb) \ga^0 \biggl(
  \int \whack A(k - k_1) P_+(\kb_1) \ga^0 \whack A(k_1 - k')
    \delta(k_1^0 - \omega(\kb_1)) \,d^4k_1 \biggr) P_+(\kb') \cr
&\quad = i(2\pi)^{-3} e^2\, P_+(\kb) \ga^0 \biggl(
  \int \whack A(k - k_1) \ga^0 S_F(k_1)^\7
   \ga^0 \whack A(k_1 - k') \,d^4k_1 \biggr) P_+(\kb'). \cr}
$$
Here we have used:
$$
\bigl\lbrack S_\ret(k)
 + 2\pi i P_+(\kb) \ga^0\delta(k^0 - \omega(\kb)) \bigr\rbrack^\7
 = \ga^0 S_F(k) \ga^0,
\eqno (8.2)
$$
where $S_F$ is the Feynman propagator:
$$
S_F(k) :=  {\whack{\,k} + m \over k^2 - m^2 + i0}.
$$

   Taking adjoints and using
$\whack A(k)^\7 = \ga^0\!\whack A(k)\ga^0$, we get:
$$
\eqalignno{
((S_{++}^{-1})^\7)_{:1} (\kb,\kb')
&= -{i\over 2\pi} e\,
    P_+(\kb) \ga^0 \whack A(k - k') P_+(\kb'), \cr
((S_{++}^{-1})^\7)_{:2} (\kb,\kb')
&= -i(2\pi)^{-3} e^2\,
    P_+(\kb) \ga^0 \biggl( \int \whack A(k - k_1) S_F(k_1)
    \whack A(k_1 - k') \,d^4k_1 \biggr) P_+(\kb'). \cr}
$$
Proceeding recursively according to~\eq{8.1}, in the same way we
obtain:
$$
\eqalignno{
((S_{++}^{-1})^\7)_{:n} (\kb,\kb')
&= -i(2\pi)^{-2n+1} e^n\, P_+(\kb) \ga^0 \biggl(
    \int \whack A(k - k_1) S_F(k_1) \whack A(k_1 - k_2) \dots \cr
&\qquad S_F(k_{n-1}) \whack A(k_{n-1} - k')
    \,d^4k_1 \dots \,d^4k_{n-1} \biggr) P_+(\kb').  \cr}
$$
The argument for the scattering of a positron is entirely
analogous.

    Similarly, for pair creation we must compute:
$$
((S_{--}^{-1})^\7 S_{+-}^\7)_{:n}
 = ((S_{--}^{-1})^\7)_{:0} (S_{+-}^\7)_{:n} + \cdots
   + ((S_{--}^{-1})^\7)_{:n-1} (S_{+-}^\7)_{:1}.
$$
Note that
$$
\eqalignno{
(S_{+-}^\7)_{:n} (\kb,\kb')
&= i(2\pi)^{-2n+1}e^n\,
   P_-(\kb)\ga^0 \int d^4k_1\cdots \int d^4k_{n-1}  \cr
&\quad\x  \whack A(k - k_1) S_\adv(k_1) \dots S_\adv(k_{n-1})
          \whack A(k_{n-1} - k') \, P_+(\kb'),  \cr}
$$
where $S_\adv(k) := (\whack{\,k} + m)/(-k^2 + m^2 + ik^0 0)$.
By using as needed \eq{8.2} under the form:
$$
S_\adv(k) - 2\pi i P_+(\kb) \ga^0 \delta(k^0 - \omega(\kb))
 = S_F(k),
$$
we can finally reexpress the whole expansion for this process in
terms of the Feynman propagators, obtaining:
$$
\eqalignno{
(S_{+-}S_{--}^{-1})_{:n} (\kb,\kb')
&= -i(2\pi)^{-2n+1} e^n\, P_+(\kb) \ga^0 \biggl(
    \int \whack A(k - k_1) S_F(k_1) \whack A(k_1 - k_2) \dots \cr
&\qquad S_F(k_{n-1}) \whack A(k_{n-1} - k')
    \,d^4k_1 \dots \,d^4k_{n-1} \biggr) P_-(\kb').  \cr}
$$

   We leave as an exercise for the reader the treatment of pair
annihilation.

   Note that the phase factor of $\vacpersamp$ has no bearing on
the probabilities for electron scattering, pair production, and
the like.

\section IX. On virtual vacuum polarization

One can compute the polarization of the vacuum, following
Ref.~\BogolShir, as the vacuum expectation value of the current:
$\<0_{\rm in}, j^\mu(x) 0_{\rm in}>$, where the current is
defined as the functional derivative
$$
j^\mu(x) := i{\bm S}^\7 {\delta{\bm S} \over \delta A_\mu(x)}.
$$
Thus, one must study the functional dependence of the phase
factor $e^{i\theta\lbrack A\rbrack}$ on the vector potential.
The information we need is encoded in the vacuum persistence
amplitude. Recall that the effective action $W$ is defined by
$\vacpersamp =: e^{iW}$. Introduce $\ga(k) = 2m^2/k^2$ and
$$
\displaylines{
G(k) := {\a\over3} \int k^{-2} (1 + \ga(k)) (1 - 2\ga(k))^{1/2}
                        \theta(1 - 2\ga(k)) \,d^4k,  \cr
G^{\mu\nu}(k) := (k^\mu k^\nu k^2 - g^{\mu\nu} k^4)\, G(k),
\cr}
$$
where $\a = e^2/4\pi$, the fine structure constant. It is easy
to see that $G(x)$ and $G^{\mu\nu}(x)$ have no support outside
the light cone. Perturbatively, to first order of approximation
we have:
$$
\eqalignno{
\abs{\vacpersamp} &\simeq \exp(-\thalf \Tr(S_{+-}S_{+-}^\7))
 \simeq  1 - \thalf \Tr(S_{+-}S_{+-}^\7)  \cr
&= 1 - \int G^{\mu\nu}(k) A_\mu(k) A_\nu^*(k) \,d^4k
 = 1 - \int G(k) \abs{j(k)}^2 \,d^4k,  \cr}
$$
after a routine calculation, where the Maxwell equations have
been used to conjure up the sources of the classical field. This
gives the imaginary part of~$W$.

   The real part of the effective action is found by means of a
dispersion relation. We gloss here the very detailed treatment
in~Ref.~\Scharf. The dispersion relation can be written
nonperturbatively in the form:
$$
{\delta \over \delta A_\nu(y)}\, {\bm S}^\7 \lbrack A\rbrack
 {\delta{\bm S} \lbrack A\rbrack \over \delta A_\mu(x)} = 0
$$
for $A$ arbitrary and $y^0 > x^0$. Adding the real and imaginary
components of the effective action, we finally arrive at:
$$
W\lbrack j\rbrack
 = {\a \over 3\pi} \int d^4k\, H(k) \abs{j(k)}^2,
$$
where
$$
H(k) = \int_{4m^2}^\infty d\la\,
 {(1 + \ga(\la)) (1 - 2\ga(\la))^{-1/2} \over \la(\la - k^2-i0)}.
$$
The latter is precisely the renormalized expression from which
one can immediately compute, for instance, the Uehling
correction to the Coulomb potential.\cite{\Dittrich}

   Now, what is the meaning of the dispersion relation? It is the
functional-differential form of the ``causality
condition'':\cite{\Scharf,\BogolShir}
$$
{\bm S}\lbrack A_1 + A_2\rbrack
 = {\bm S}\lbrack A_2\rbrack\, {\bm S}\lbrack A_1\rbrack
$$
when $A_2$ is to the future of $A_1$. This is nothing but our
cocycle condition~\eq{4.8} for ortho\-gonal transformations
parametrized by the gauge potential:
$$
e^{i\theta\lbrack A_1 + A_2\rbrack}
 = e^{i\theta\lbrack A_1\rbrack} e^{i\theta\lbrack A_2\rbrack}\,
 \exp\bigl( i\arg \det(I + S_{--}^{-1}\lbrack A_2\rbrack
   S_{-+}\lbrack A_2\rbrack S_{+-}\lbrack A_1\rbrack
   S_{--}^{-1}\lbrack A_1\rbrack) \bigl).
$$
In the last formula the equation~\eq{7.8} has been used.

\bigskip\bigskip

\leftline{\bf Acknowledgements}

\bigskip

We are grateful for helpful discussions with J.~A.~Dixon,
S.~A.~Fulling, G.~Moreno, P.~Morley, J.~Polchinski and
E.~C.~G.~Sudarshan. We benefitted from visits to the
International Centre for Theoretical Physics, Trieste and the
Universidad de Zaragoza (JCV), and the Forschungszentrum BiBoS
of the Universit\"at Bielefeld and the University of Texas at
Austin (JMGB). We also thank the referee, whose comments
helped to improve the presentation of the paper.

\bigskip\bigskip

\leftline{\bf References}

\bigskip
\frenchspacing
\parindent=10pt

\refno\curraddr. Permanent address: Escuela de Matem\'atica,
Universidad de Costa Rica, San Jos\'e, Costa Rica.

\smallskip

\refno\ShaleStine. D. Shale and W. F. Stinespring,
J. Math. Mech. {\bf 14}, 315 (1965).

\refno\Pressley. A. Pressley and G. Segal,
{\it Loop Groups\/} (Clarendon Press, Oxford, 1986).

\refno\Mickelsson. J. Mickelsson,
{\it Current algebras and groups\/} (Plenum, New York, 1989).

\refno\Connes. A. Connes,
Publ. Math. IHES {\bf 62}, 257 (1985).

\refno\Araki. H. Araki,
Contemp. Math. {\bf 62}, 23 (1987).

\refno\Bourbaki. N. Bourbaki,
{\it Alg\`ebre\/} (Hermann, Paris, 1959), Chap. 9.

\refno\Emch. G. G. Emch,
{\it Algebraic Methods in Statistical Mechanics
and Quantum Field Theory\/} (Wiley, New York, 1972).

\refno\Taylor. M. E. Taylor,
{\it Noncommutative Harmonic Analysis}, Math. Surveys and
Monographs {\bf22} (Am. Math. Soc., Providence, RI, 1986).

\refno\Segalcwave. I. E. Segal,
Adv. Math. Suppl. Studies {\bf 3}, 321 (1978).

\refno\Ruijsenaars. S. N. M. Ruijsenaars,
Ann. Phys. (N.Y.) {\bf 116}, 105 (1978).

\refno\Lawson. H. B. Lawson and M. L. Michelsohn,
{\it Spin Geometry\/} (Princeton U. P., Princeton, 1989).

\refno\Sirius. J. C. V\'arilly and J. M. Gracia-Bond{\'\i}a,
J. Geom. Phys. {\bf 12}, 1 (1993).

\refno\CareyPalmer. A. Carey and J. Palmer,
J. Func. Anal. {\bf 83}, 1 (1989).

\refno\Palmer. J. Palmer,
J. Math. Phys. {\bf 29}, 1283 (1988).

\refno\Popov. A. D. Popov and A. G. Sergeev,
``Bosonic strings, ghosts and geometric quantization'',
preprint JINR--E2--92--261, Dubna, 1992.

\refno\Borthwick. D. Borthwick, S. Klimek, A. Lesniewski and M.
Rinaldi, Commun. Math. Phys. {\bf 153}, 49 (1993).

\refno\GrosseMad. H. Grosse, W. Maderner and C. Reitberger,
J. Math. Phys. {\bf 34}, 4409 (1993).

\refno\Mickelssonbis. J. Mickelsson,
Commun. Math. Phys. {\bf 154}, 403 (1993).

\refno\Maderner. W. Maderner,
J. Phys. A {\bf 25}, 2489 (1992).

\refno\Weinberg. S. Weinberg,
Phys. Rev. B {\bf 133}, 1318 (1964).

\refno\Bongaarts. P. J. M. Bongaarts,
Ann. Phys. (N.Y.) {\bf 56}, 108 (1970).

\refno\Saunders. S. Saunders,
in {\it The Philosophy of Vacuum}, S. Saunders and H. R. Brown,
eds. (Oxford U. P., Oxford, 1991).

\refno\Thaller. B. Thaller,
{\it The Dirac Equation\/} (Springer, Berlin, 1992).

\refno\Careyob. A. L. Carey and D. M. O'Brien,
Lett. Math. Phys. {\bf 6}, 335 (1982).

\refno\Matsui. T. Matsui,
J. Func. Anal. {\bf 94}, 93 (1990).

\refno\AGaume. L. Alvarez-Gaum\'e and M. A. V\'azquez-Mozo,
``Topics in String Theory and Quantum Gravity'', lectures
delivered at Les Houches Summer School, FTUAM--92/38 \&
CERN-TH-6736/92 preprint, Geneva, 1992.

\refno\Ruijsenaarsbis. S. N. M. Ruijsenaars,
Commun. Math. Phys. {\bf 124}, 553 (1989).

\refno\Aldebaran. J. C. V\'arilly and J. M. Gracia-Bond{\'\i}a,
Mod. Phys. Lett. A {\bf 7}, 659 (1992).

\refno\Scharf. G. Scharf,
{\it Finite Quantum Electrodynamics\/} (Springer, Berlin, 1989).

\refno\Feynman. R. P. Feynman,
Phys. Rev. {\bf 76}, 749 (1949).

\refno\BogolShir. N. N. Bogoliubov and D. V. Shirkov,
{\it Introduction to the Theory of Quantized Fields}, 3rd edition
(Wiley, New York, 1980).

\refno\Dittrich. W. Dittrich and M. Reuter,
{\it Effective Lagrangians in Quantum
Electrodynamics\/}\hfil\break
(Springer, Berlin, 1985).

\bye